\documentclass[a4paper,11pt]{article}
\usepackage{jcappub} 
\usepackage{multirow}
\usepackage{color, xcolor}

\arxivnumber{2406.04407} 
\title{\boldmath Reconstructing redshift distributions with photometric galaxy clustering}

\author{Hui Peng}
\author[1]{and Yu Yu\note{Corresponding author.}}
\affiliation{Department of Astronomy, School of Physics and Astronomy, Shanghai Jiao Tong University, Shanghai 200240, China}
\affiliation{Key Laboratory for Particle Astrophysics and Cosmology (MOE)/Shanghai Key Laboratory for Particle Physics and Cosmology, Shanghai 200240, China}

\emailAdd{pengh.me@gmail.com, yuyu22@sjtu.edu.cn}

\abstract{The accurate determination of the true redshift distributions in tomographic bins is critical for cosmological constraints from photometric surveys. The proposed redshift self-calibration method, which utilizes the photometric galaxy clustering alone, is highly convenient and avoids the challenges from incomplete or unrepresentative spectroscopic samples in external calibration. However, the imperfection of the theoretical approximation on broad bins as well as the flaw of the algorithm in previous work 
\cite{Peng:2022aa} 
risk the accuracy and application of the method. In this paper, we propose the improved self-calibration algorithm that incorporates novel update rules, which effectively accounts for heteroskedastic weights and noisy data with negative values. The improved algorithm greatly expands the application range of self-calibration method and accurately reconstructs the redshift distributions for various mock data. Using the luminous red galaxy (LRG) sample of the Dark Energy Spectroscopic Instrument (DESI) survey, we find that the reconstructed results are comparable to the state-of-the-art external calibration.
This suggests the exciting prospect of using photometric galaxy clustering to reconstruct redshift distributions in the cosmological analysis of survey data.}

\begin{document}
\maketitle
\flushbottom

\section{Introduction}\label{sec:intro}

The ongoing and upcoming photometric (imaging) galaxy surveys, such as the Hyper Suprime-Cam survey (HSC; \cite{Aihara:2018tn}), the Kilo-Degree Survey (KIDS; \cite{Kuijken:2019vi}), the Dark Energy Survey (DES; \cite{Abbott:2018aa}), the Dark Energy Spectroscopic Instrument (DESI; \cite{DESI-Collaboration:2016vs,DESI-Collaboration:2016vy}) Legacy Imaging Surveys, Euclid \cite{Euclid-Collaboration:2024aa}, the China Space Station Telescope (CSST; \cite{Gong:2019tb}) and the Legacy Survey of Space and Time (LSST; \cite{The-LSST-Dark-Energy-Science-Collaboration:2018vo}), provide strong constraints on cosmological parameters using the clustering of galaxies and weak gravitational lensing.
Nevertheless, the cosmological estimates derived from extensive photometric samples of galaxies are approaching a threshold beyond which expanding the sample size alone is unlikely to yield a marked enhancement in the precision of the constraints \cite{Becker:2016aa,Hildebrandt:2021vt,Myles:2021aa}.
Quantifying and mitigating systematic errors is becoming crucial for reducing uncertainty in cosmological analysis \cite{Mandelbaum:2018aa,Yao:2024aa}.
The major one challenge is the uncertainties of redshift distributions due to limited accuracy in the estimated photometric redshifts \cite{Bernstein:2010wc,Salvato:2019aa,Newman:2022vy}.

Several redshift calibration techniques utilizing spectroscopic sample have been proposed in the last decades, which can be broadly separated in three categories:
(i) weighting the redshift distribution of reference galaxy samples, such as direct redshift calibration (DIR; \cite{Lima:2008vl,Bonnett:2016wb,Hildebrandt:2017ua,Wright:2019aa,Hildebrandt:2020aa}) and self-organizing map (SOM; \cite{Kohonen1982}) based schemes \cite{Masters:2015wp,Herbel:2017aa,Buchs:2019tx, Wright:2020vs,Hildebrandt:2021vt,Myles:2021aa,Giannini:2024aa};
(ii) cross-correlating with reference spectroscopic galaxy samples \cite{Newman:2008vv,Matthews:2010un,McQuinn:2013uh,Schmidt:2013va,Choi:2016aa,Hildebrandt:2017ua,McLeod:2017vz,Morrison:2017aa,Gatti:2018tg,Hildebrandt:2020aa,Hildebrandt:2021vt,Myles:2021aa,Cawthon:2022aa,Gatti:2022va,Rau:2022ud,Stolzner:2023aa,Giannini:2024aa};
(iii) Bayesian hierarchical approaches \cite{Leistedt:2016aa,Leistedt:2019aa,Sanchez:2019aa,Alarcon:2020aa,Malz:2022aa,Autenrieth:2024aa}.
Despite significant advancements and extensive application, these external calibration methods are typically restricted by a number of effects from spectroscopic samples like incompleteness and sample variance, leading to potential errors that could surpass what is acceptable for precise cosmological constraints \cite{Gruen:2017aa,Hemmati:2019aa,Hartley:2020aa,Sanchez:2020aa,Myles:2021aa,Newman:2022vy}.
Substantial advancements in redshift calibration are essential for the success of both current and forthcoming photometric surveys.

In addition to the mentioned approaches above, self-calibration techniques completely independent of spectroscopic data have been suggested and developed \cite{Schneider:2006ta,Benjamin:2010aa,Zhang:2010wr,Zhang:2017um,Schaan:2020up,Pyne:2021aa,Stolzner:2021va,Peng:2022aa,Xu:2023aa,Song:2024aa}.
This approach possesses significant potential, particularly for future combined analysis with calibration methods using spectroscopic data, owing to its ability to derive constrains from the photometric data alone.
Besides, a forward modeling framework recently proposed in \cite{Alsing:2023aa,Leistedt:2023aa}, establishes a direct link between photometric redshift inference and the physics of galaxy evolution.
This approach holds the potential to address numerous challenges, yet it remains in the early stages of development and require further improvements in modeling and optimization of the model hyper-parameters.

The self-calibration algorithm in \cite{Zhang:2017um,Peng:2022aa,Xu:2023aa}, is extremely convenient to apply, allowing for the direct input of galaxy–galaxy correlations between photometric redshift bins without the need for any prior or complex preliminary steps.
The theoretical basis is that the galaxy-galaxy correlation signals between different photometric redshift bins originate from the combination of galaxy auto correlations in true redshift bins, due to photometric redshift uncertainties.
Consequently, these non-zero galaxy-galaxy correlations between photometric redshift bins can be utilized to infer the redshift distributions.
The algorithm has been deployed to reconstruct the redshift scatter rates as well as the auto power spectra in true redshift bins, reduce the mean redshift bias, and measure the photometric Baryonic Acoustic Oscillations (BAO), among other applications \cite{Peng:2022aa,Xu:2023aa,Song:2024aa}.
These studies typically utilize broad redshift bins to mitigate the impact of noisy data characterized by a low signal-to-noise ratio (SNR).
Nevertheless, the effectiveness of the theory underlying the self-calibration algorithm may diminish when applied to broad tomographic bins with realistic redshift distributions, an issue that has been both neglected and insufficiently addressed in previous studies.
The disregard for the inherent limitation of theoretical approximation, coupled with the potential shortcomings in algorithm design, may pose a non-negligible threat to the accuracy and overall applicability of self-calibration method in various contexts.
It is imperative to implement optimization efforts to refine the effectiveness of the current approach.

In this work, we address this challenge by introducing an improved algorithm with new update rules in the iterative process of non-negative matrix factorization (NMF), and adequately handling the nonuniform uncertainties and noisy data with negative values.
The enhanced algorithm significantly ensures the stability and accuracy of the reconstructions over thin redshift bins, which are essential for holding the self-calibration method.
Furthermore, with a shift towards thinner bins, the reconstructed results can be more effectively characterize the detailed redshift distributions.

The paper is organized as follows.
Section~\ref{sec:method} introduces the self-calibration theoretical framework and the novel update rules to improve the algorithm.
Section~\ref{sec:data} describes all the sample data used in this work, both from survey and mock.
The modified algorithm is tested on mock data and subsequently validated using DESI luminous red galaxies in section~\ref{sec:results}.
Finally, in section~\ref{sec:dis_cons}, we conclude with a discussion of our findings and discuss their implications.

\section{Methodology}\label{sec:method}
\subsection{Theoretical framework}\label{sec:theory}
The presence of substantial uncertainties in photometric redshift measurements results in a considerable fraction of galaxies within a photometric redshift bin coming from various true redshifts.
Assume that we split galaxies into $n$ tomographic redshift bins.
We denote the ratio of the galaxies in true redshift (true-$z$) bin $i$ but observed in photometric redshift (photo-$z$) bin $j$ as $P_{ij}\,{\equiv}\,N_{i{\rightarrow}j}/N_j^P$.
Here, $N_{i{\rightarrow}j}$ represents the count of galaxies that are misclassified from the true-$z$ bin $i$ to the photo-$z$ bin $j$, and $N_j^P$ is the total number of galaxies within the photo-$z$ bin $j$.
For each photo-$z$ bin $j$, there exists the normalization $\sum_iP_{ij}=1$.
We note that the $P_{ij}$ scatter for each photo-$z$ bin $j$ is equivalent in meaning to the redshift distribution from true-$z$ bins, and thinner bins provide a more detailed distribution.

The galaxy-galaxy angular power spectrum of two photo-$z$ bins $C_{ij}^{gg,P}$ and true-$z$ bins $C_{ij}^{gg,R}$ can be connected through the following relationship:
\begin{equation}
    C_{ij}^{gg,P}(\ell)=\sum_kP_{ki}P_{kj}C_{kk}^{gg,R}(\ell)+\delta{N_{ij}^{gg,P}(\ell)}\ .
    \label{eqn:CggPsum}
\end{equation}
This equation has ignored $C_{k{\neq}m}^{gg,R}$, since the galaxy cross-correlation between non-overlapping redshift bins is sufficiently weak to be considered negligible.
The last term $\delta{N_{ij}^{gg,P}}$ is the associated shot noise fluctuation after the subtraction of its ensemble average.
The fluctuation level can be well approximated by a Gaussian distribution with
\begin{equation}
    \sigma_{i j}^{g g, P}=\sqrt{\frac{1}{(2 \ell+1) \Delta \ell f_{\text{sky}}} \frac{1+\delta_{i j}}{\bar{n}_{i} \bar{n}_{j}}}\ , 
    \label{eqn:noise}
\end{equation}
where $\Delta\ell$ is the $\ell$ bin size, $f_\mathrm{sky}$ is the sky fraction, $\bar{n}_i$ ($\bar{n}_j$) is the number density in photo-$z$ bin $i$ ($j$), and $\delta_{i j}$ is the Kronecker Delta.
For a given $\ell$, we can rewrite the eq.~(\ref{eqn:CggPsum}) in matrix form,
\begin{equation}
    C_{\ell}^{gg,P}=P^{T}C_{\ell}^{gg,R}P+\delta{N_{\ell}^{gg,P}}\ .
    \label{eqn:CP}
\end{equation}
We denote the matrix $P$ as scattering matrix.
Besides, a similar relationship exists between the lensing convergence and galaxy number density \cite{Zhang:2010wr}, which, however, has not yet been integrated into the existing algorithm.

Previous studies (e.g. \cite{Schneider:2006ta,Zhang:2010wr,Zhang:2017um}) make a crucial assumption: the photometric galaxy clustering signal can be accurately obtained from the galaxy clustering in true redshift bins by using eq.~(\ref{eqn:CP}).
This assumption only holds if the galaxies scattered into a given photo-$z$ bin are representative subsamples of those in the corresponding true-$z$ bins.
The simulated data in \cite{Peng:2022aa} satisfy this condition, as the galaxies in each photo-$z$ bin are randomly selected from the true-$z$ bins according to the preset fractions.
However, galaxies from true-$z$ bins are unlikely to be a fully random subsample due to the specific characteristics of the redshift distribution profile, such as a Gaussian-like shape.
The impact of this weakening becomes particularly significant in the context of broad redshift bins that were commonly used in past analyses.

To address this issue and expand the applicability of the self-calibration method, it is recommended to utilize the thinnest possible bin.
However, the reconstruction requirement for thin tomographic bins introduces several key challenges: (i) an increase in noise level due to reduced galaxy number density within each bin; (ii) the necessity to fit a greater number of parameters, leading to higher degeneracy; and (iii) a rise in scattering dispersion, with the diagonal elements of the scattering matrix no longer being dominant.
These complexities introduce substantial difficulties for existing algorithms.
In the subsequent sections, we will discuss the limitations of current self-calibration algorithm and the advancements we have implemented.

\subsection{Non-negative matrix factorization}\label{sec:nmf}
The original algorithm for photometric redshift self-calibration is composed of two main components: (i) a fixed-point iteration algorithm, referred to as Algorithm 1, designed to provide an initial set of plausible results; and (ii) an iteration based on NMF, known as Algorithm 2, which further refines the output of Algorithm 1 to find the optimal solution.
The self-calibration algorithm aims to find reliable results through the minimization of objective function
\begin{equation}
    \mathcal{J} = \frac{1}{2} \sum_{\ell}\left\|C_{\ell}^{g g, P}-P^{T} C_{\ell}^{g g, R} P\right\|_{F}^{2}\ ,
    \label{define:J}
\end{equation}
where $\|.\|_F$ is the Frobenius form.
$\mathcal{J}$ measures the accumulation of decomposition error across all data matrices between the observations and reconstructions.
This section avoids elaborating on the entire algorithm, and the readers are encouraged to consult the referenced papers which describe the full details in \cite{Zhang:2017um,Peng:2022aa,Xu:2023aa}.

Our focus lies on the update rules within the algorithm that employs the NMF technique, which serves as the core of the reconstruction process.
NMF is a dimensionality reduction technique that has shown promise in various fields \cite{Paatero:1994aa,Lee:2001aa}.
Astrophysics is at the forefront of investigating novel applications of this technique, especially exploring its potential implications on data processing \cite{Tsalmantza:2012aa,Zhu:2016aa,Ren:2018aa,Green:2023aa}.
We provide a concise introduction of the NMF framework, which is, for a given matrix $X$, find matrix factors $W$ and $H$ such that
\begin{equation}
    X \approx W H\ ,
    \label{nmf}
\end{equation}
with the constraints $X, W, H \in \mathbb{R}_{+}$, i.e. all values in all matrices are non-negative.
We follow \cite{Zhang:2017um} and use the alternative notations $X \equiv C_{\ell}^{g g, P}, W \equiv P^T$ and $H \equiv C_{\ell}^{g g, R} P$ to clearly describe the technique without loss of generality.
And then we have the row-sum-to-one constrain: $\sum_jW_{ij}=1$.

To find the two factors $W$ and $H$, the goal of NMF is to minimize the objective function
\begin{equation}
    \mathcal{J}=\|X-W H\|_{F}^2\ ,
    \label{nmf_obj}
\end{equation}
which is similar to eq.~(\ref{define:J}).
Suppose we are minimizing an objective function over matrix $W$, the multiplicative update is easily derived from the gradient $\nabla=\partial \mathcal{J} / \partial W$.
The conventional multiplicative update rule can be given by
\begin{equation}
    W \leftarrow W \circ \frac{\nabla^{-}}{\nabla^{+}} = W \circ \frac{X H^T}{W H H^T}\ ,
    \label{eq:w_update_0}
\end{equation}
where the open circle dot operator $\circ$ denotes element-wise (Hadamard) multiplication, $\nabla^{+}$ and $\nabla^{-}$ are the positive and (unsigned) negative parts of the gradient, respectively, i.e. $\nabla=\nabla^{+}-\nabla^{-}$.
Ref.~\cite{Zhang:2017um} use the iterative Lagrangian solution proposed in \cite{Zhu:2013aa} to solve the non-negative learning problem with stochasticity constraints, i.e. $\sum_jW_{ij}=1$ here.
The objective function is nonincreasing by using the following multiplicative update rule:
\begin{equation}
    W \leftarrow W \circ \frac{\nabla^{-}\circ A+1-B}{\nabla^{+}\circ A}\ ,
    \label{w_update_zhang17_1}
\end{equation}
where $\nabla^{-}=\sum_{\ell}\left[V_{\ell} H_{\ell}^T\right]$, $ \nabla^{+}=\sum_{\ell}\left[W H_{\ell} H_{\ell}^T\right]$, $A_{i j}=\sum_b W_{i b} / \nabla_{i b}^{+}$ and $B_{i j}=\sum_b W_{i b} \nabla_{i b}^{-} / \nabla_{i b}^{+}$.
The sum of $\ell$ components is derived from the requirement of the objective function in eq.~(\ref{define:J}).
Besides, the “moving term” trick \cite{Zhu:2010aa} is applied to overcome negative entries may caused by $-B$ in the numerator, which gives
\begin{equation}
    W \leftarrow W \circ \frac{\nabla^{-}\circ A+1}{\nabla^{+}\circ A+B}\ .
    \label{w_update_zhang17_2}
\end{equation}

We note that additional algorithmic details like update rules for $C_{\ell}^{g g, R}$ are omitted here.
The algorithm, with the update rule above, has shown promising results when handling data with high SNR, given that the eq.~(\ref{eqn:CP}) is satisfied.
However, as mentioned in section~\ref{sec:theory}, the application of this approach to realistic scenarios demands thinner bin resolutions than the current algorithm can accommodate (see appendix~\ref{sec:appendix_compare_algorithm} for more details).

\subsection{Novel update rules for NMF}\label{sec:new update rules}
It is obvious that the NMF technique as presented above does not account for two critical challenges:
\vspace{1em}

(i) the objective function and the associated update rules are based on the assumption of homoscedastic data, but in astronomical applications, we frequently encounter nonuniform uncertainties and missing data.

(ii) there is a lack of statistically consistent treatment for negative data values, which is particularly problematic for low signal-to-noise data with many negative values.
\vspace{1em}

These deficiencies can significantly impair the utility of self-calibration algorithm.
Fortunately, we find that the effects of these challenges have been studied and are essentially solved in \cite{Green:2023aa}.
But their update rules do not include the additional constraint like $\sum_jW_{ij}=1$ required for the scenarios we are addressing.
Our contributions in this paper amount to integrating update rules for NMF that can correctly account for nonuniform uncertainties and negative data elements, while preserving the exceptional performance of the original algorithm.

The Nearly-NMF technique presented in \cite{Green:2023aa}, based on weighted vectorized update rules derived in \cite{Zhu:2016aa} for heteroscedastic non-negative data, provides the capacity to adequately handle noisy data with negative values.
We provide a brief summary here and refer readers to \cite{Green:2023aa} for details of the methodology.
The objective function for this innovative NMF technique is 
\begin{equation}
    \mathcal{J}=\left\|V^{1 / 2} \circ\left((X+Y)-(W H+Y)\right)\right\|_{F}^2\ ,
    \label{obj_fun_green}
\end{equation}
where $Y$ is a shift matrix. The weight $V$ has the same dimension as $X$ and a common choice is the inverse variance matrix of each matrix element $1/\sigma^2$.
Similar to eq.~(\ref{eq:w_update_0}), the update rule can be derived and given by
\begin{equation}
    W \leftarrow W \circ \frac{(V \circ(X+Y)) H^T}{(V \circ(W H+Y)) H^T}\ .\label{non_nmf}
\end{equation}

The matrix $Y$ in Nearly-NMF technique is determined as the absolute minimum possible shift, $Y_{\rm min}$ for quickest convergence.
Since $V, W, H, Y \in \mathbb{R}_{+}$, the minimum shift $Y_{\rm min}$ is defined by only requiring
\begin{equation}
    (V \circ(X+Y)) H^T\geq 0 \ ,
    \label{nearlynmf_require}
\end{equation}
and we can derive the minimum shift by
\begin{equation}
    (V \circ Y_{\rm min}) H^T=\left[(V \circ X) H^T\right]^{-} .
    \label{minimum_shift}
\end{equation}
The notation $[M]^{-}$ refers to a matrix containing the absolute value of the negative elements of $M$, with the formerly positive values set to zero. Correspondingly the notation $[M]^{+}=M+[M]^{-}$.
Rewriting update rule~(\ref{non_nmf}) with these notations, the final update rule for Nearly-NMF is given by
\begin{equation}
    W \leftarrow W \circ \frac{\left[(V \circ X) H^T\right]^{+}}{(V \circ(W H)) H^T+\left[(V \circ X) H^T\right]^{-}}\ .
    \label{w_update_green}
\end{equation}

For our application, the objective function with heteroskedastic weights $V$ is
\begin{equation}
    \mathcal{J} = \frac{1}{2} \sum_{\ell}\left\|V^{1 / 2} \circ \left(X-WH\right)\right\|_{F}^{2}\ .
    \label{define:J_new}
\end{equation}
A natural and straightforward combination idea is to consider the numerator and denominator in update rule (\ref{w_update_green}) as new $\nabla^{-}$ and $\nabla^{+}$ in previous update rule (\ref{w_update_zhang17_1}), respectively.
This substitution produces our final multiplicative update rule:
\begin{align}
    &\ W \leftarrow W \circ \frac{\nabla^{-}\circ A+1-B}{\nabla^{+}\circ A}\ , \\
    &\ \nabla^{-} = \sum_{\ell}\left\{\left[(V \circ X) H^T \right]^{+}\right\}\ ,\notag \\
    &\ \nabla^{+} =\sum_{\ell}\left\{(V \circ(W H)) H^T+\left[(V \circ X) H^T\right]^{-}\right\}\ ,\notag
    \label{w_update_final}
\end{align}
where $A_{i j} =\sum_b W_{i b} / \nabla_{i b}^{+}$, $B_{i j} =\sum_b W_{i b} \nabla_{i b}^{-} / \nabla_{i b}^{+}$.
The same substitution operation is applied to update rule (\ref{w_update_zhang17_2}) in case of any negative entries in the updated $W$.
We have integrated this novel update rule and objective function into the self-calibration algorithm, replacing those previously utilized in the NMF iterative process, and this modification has markedly improved the robustness and effectiveness.

\section{Data}\label{sec:data}
\subsection{DESI LRG sample}\label{sec:LRG_data}
\begin{figure}[htbp]
    \centering
	\includegraphics[width=10cm]{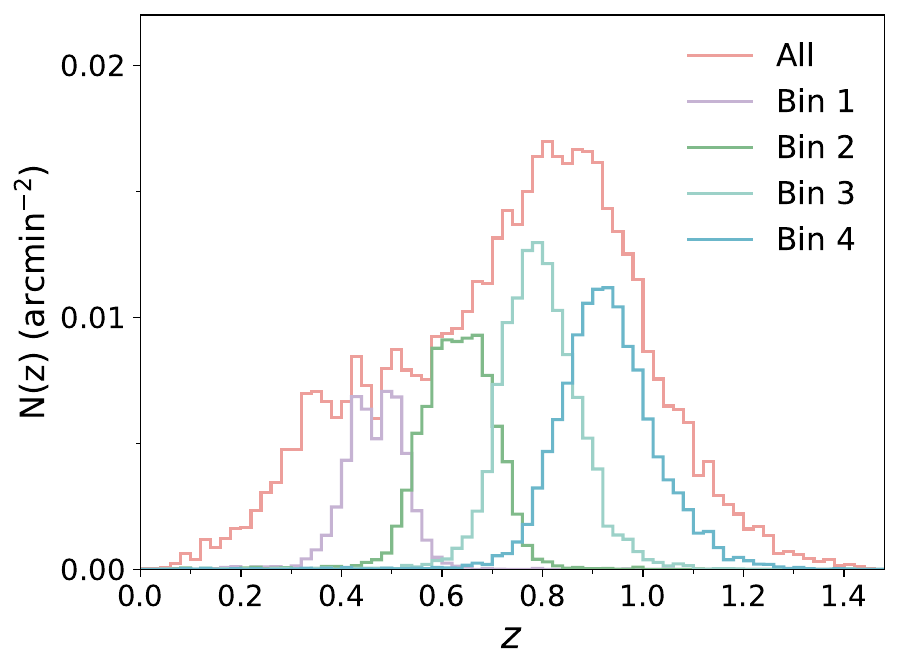}
    \caption{The redshift distributions ($0<z<1.46$) of the full DESI LRG sample and the four subsamples in \cite{Zhou:2023aa}, based on 22k DESI spectroscopic redshifts. Units of the y-axis are the number of galaxies per square arcmin within the redshift bin with width $dz = 0.02$.}
    \label{fig:n_z}
\end{figure}
The DESI experiment is the first on-going Stage IV galaxy survey, designed to conduct a five-year survey with about 40 million extra-galactic spectroscopic redshifts over 14,000~$\mathrm{deg}^2$ coverage \cite{DESI-Collaboration:2016vs,DESI-Collaboration:2016vy,DESI-Collaboration:2022aa,DESI-Collaboration:2024ab}.
DESI uses an optimized tracer for each targeted redshift range \cite{Myers:2023aa}.
The selections based on optical and infrared imaging data are adopted to identify a bright sample of low redshift galaxies (BGS; \cite{Hahn:2023ab}) covering $0.05 < z < 0.4$, luminous red galaxies (LRG; \cite{Zhou:2023ab}) covering $0.4<z\lesssim 1.0$, emission line galaxies (ELG; \cite{Raichoor:2023aa}) covering $0.6<z<1.6$, quasars (QSO; \cite{Chaussidon:2023aa}) as direct tracers $(0.9<z<2.1)$, and Lyman-$\alpha$ forest quasars ($z>2.1$) as a probe of the intergalactic medium.

In this work, we use a public extended LRG sample\footnote{\href{https://data.desi.lbl.gov/public/papers/c3/lrg_xcorr_2023/}{https://data.desi.lbl.gov/public/papers/c3/lrg\_xcorr\_2023/}} from \cite{Zhou:2023aa}, with 2-3 times the DESI Main Survey LRG target sample density described in \cite{Zhou:2023ab}.
This sample is selected from the DESI Legacy Imaging Surveys (LS) Data Release 9 \cite{Dey:2019aa}.
After applying the masks and footprint trimming, the total area is $\sim$$16,700~\mathrm{deg}^2$, with surface density $\sim$$0.46~\mathrm{arcmin}^{-2}$.
The photo-$z$ ($z_p$) performance of the extended LRG sample is assessed by using the DESI spectroscopic redshifts ($z_s$).
The overall photo-$z$ error, quantified by the normalized median absolute deviation defined as $1.48 \times \operatorname{median}(|\Delta z| /(1+z_s))$ where $\Delta z=z_p-z_s$, is 0.027 (0.031) in South (North), and the outlier fraction, defined as having $|\Delta z|>0.1 \times(1+z_s)$, is $2.1 \%$ (3.2\%) in South (North).

Although the majority of this LRG sample will not be observed by the DESI spectroscopy, a large number spectra within have already obtained from DESI Survey Validation (SV1; \cite{DESI-Collaboration:2024ad,DESI-Collaboration:2024aa}) which are adequate for characterizing their redshift distributions.
Actually, this sample is a subset of the SV1 IR sample, and IR is one of two selections for SV1 LRG sample \cite{Zhou:2023ab}.
Therefore, ref.~\cite{Zhou:2023aa} use 22k spectroscopic redshifts of this LRG sample from the DESI SV1 data, which covering roughly $230~\mathrm{deg}^2$ ($100~\mathrm{deg}^2$ in the North and $130~\mathrm{deg}^2$ in the South), to obtain the redshift distributions of full sample and four subsamples (Bins 1-4) shown in figure~\ref{fig:n_z}.
We note that the original four subsamples are obtained using slightly different redshift selections from North and South (see Table 3 of \cite{Zhou:2023aa}).
Considering the redshift distribution resolution of 0.02, and to facilitate the comparison of our reconstructed results later, we define and select these four subsamples with edges $z_p=$ 0.40, 0.54, 0.72, 0.86, and 1.0.
Even though the selected samples may exhibit minor differences from the original, these discrepancies do not meaningfully affect the analysis or the main conclusions.

We consider the redshift distributions obtained by \cite{Zhou:2023aa} to be state-of-the-art and utilize them as benchmark references in this paper.
Nevertheless, it is essential to acknowledge the inherent uncertainties that arises from the limited area and small number of spectroscopic redshifts available.

\subsection{Mock data}\label{sec:mock_data}
To evaluate the performance of the enhanced algorithm featuring new updated rules, we generate various mock datasets using the reference redshift distribution derived from the DESI LRG sample.
We select several specific analysis configurations.
Firstly, the photometric redshifts of mocks are obtained by adding the photo-$z$ errors, which have multiple components as 1\% uniform distribution, 4\% Gaussian distribution with $\sigma_z=0.1(1+z_s)$, and 95\% Gaussian distribution with (i) $\sigma_z=0.03(1+z_s)$ (hereafter Low Photo-$z$ Error) or (ii) $\sigma_z=0.05(1+z_s)$ (hereafter High Photo-$z$ Error).
Besides, two distinct schemes are implemented to model and manage the shot noise level with eq.~(\ref{eqn:noise}) by varying the area and surface density: (i) one scheme aligns with the DESI LRG sample above (hereafter High Noise); and (ii) the other scheme is similar to \cite{Xu:2023ab}, designed to have a lower shot noise (hereafter Low Noise), covers a smaller area but features a higher surface density, with values of $10,000~\mathrm{deg}^2$ and $2~\mathrm{arcmin}^{-2}$, respectively.
Note that the noise levels in the two scenarios discussed are significantly higher than what is typically expected for stage-IV cosmic shear samples (e.g. LSST) whose noise should be much smaller due to high galaxy number density.

Applying the Limber approximation, the galaxy-galaxy angular power spectrum between two tomographic bins $i$ and $j$ is given by
\begin{equation}
C_{ij}^{gg}(\ell)= \int_0^{\chi_{\mathrm{H}}} \frac{d \chi}{\chi^2} K_i(\chi) K_j(\chi) P_{\delta}\left(k=\frac{\ell+1 / 2}{\chi}, z\right)\ ,
\end{equation}
with $K(\chi)=H(z) b_g(z) n(z)$.
$\chi$ and $\chi_{\mathrm{H}}$ is the comoving radial distance and the co-moving horizon distance, respectively.
$H(z)=dz/d\chi$ is the expansion rate at redshift $z$, $b_g(z)$ is the linear galaxy bias, and $n(z)$ is the redshift distribution within tomographic bin.
$P_{\delta}$ is the matter power spectrum.
We assume the linear galaxy biases follow $b_g(z)=1.5/D(z)$, where $D(z)$ is the linear growth factor normalized by $D(z = 0)=1$.
The idealized model of galaxy bias is scale-independent, linear and deterministic, which may not fully capture the complexities of practical scenarios.
We do not take into account the possible unique bias properties of outlier galaxies, which may significantly differ from the core population, such as those originating from particular galaxy types.
Neglecting this difference in galaxy bias could potentially have non-negligible implications for the self-calibration method.
Nevertheless, in the mock data construction, inside a given true-$z$ bin the redshift distribution profiles for each photometric redshift bin are different, indicating that the outlier galaxies from the true-$z$ bins are not a random subsample, and at least their mean redshifts are generally different.
Therefore, although there are some simplifications, the galaxy biases of outlier galaxies still exhibit some distinctions from those of the main galaxy population within each true-$z$ bin.
We adopt a fiducial $\Lambda$CDM cosmology with parameter values from the Planck 2018 TT,TE,EE+lowE+lensing+BAO results \cite{Planck-Collaboration:2020aa}.

One key difference from previous works is that, we directly use the galaxy-galaxy power spectra in photometric redshift bins with redshift distributions characterized by realistic photo-$z$ errors.
This approach contrasts with the aforementioned potentially risky simplifications, such as using the synthetic data directly generated by the right-hand side of eq.~(\ref{eqn:CP}), or randomly sampling a subset of simulated galaxies from true-$z$ bins.
The redshift range for all mock data tests is constrained between 0.1 and 1.3, given that too low number density outside this range, making it meaningless for quantitative analysis.
However, for the application to the DESI LRG sample, we utilize two broad bins on both redshift edges to encompass the complete redshift range ($0 < z < 1.46$) in \cite{Zhou:2023aa}.

\section{Results}\label{sec:results}
In this section, we present the reconstruction results of self-calibration algorithm from mock data and DESI LRG sample.
Throughout all the analysis, we utilize the $\ell$ modes from $\ell_{\rm{min}}=100$ to $\ell_{\rm{max}}=1000$ and divide into 6 broad bands with edges $\ell=$ 100, 418, 583, 711, 819, 914 and 1000.
For the mock datasets, the \texttt{CCL}\footnote{\href{https://github.com/LSSTDESC/CCL}{https://github.com/LSSTDESC/CCL}} from \cite{Chisari:2019wv} is employed to calculate the angular power spectra.
Note that the Limber approximation used on $\ell>100$ achieves the modelling accuracy needed for Stage IV surveys \cite{Kilbinger:2017aa,Leonard:2023aa}.
For DESI LRG sample, we use the HEALPix \cite{Gorski:2005te} with $N_{\text{side}}=1024$ to construct galaxy overdensity map, and measure the angular power spectrum of galaxies with the function \textsc{compute\_coupled\_cell} in \texttt{namaster} \cite{Alonso:2019aa}.
When implementing the self-calibration algorithm, we traditionally input $\ell C_{\ell}^{gg,P}$ following the previous work.
Nevertheless, for this modified version, multiplying by $\ell$ is no longer necessarily required, as the weighting has already been integrated into the new update rules for NMF.
For the mock datasets in section~\ref{sec:results_mock}, we apply this algorithm to 100 realizations, each with an independent Gaussian random noise generated using eq.~(\ref{eqn:noise}) and identical signal power spectra.
The same perturbation procedure is also conducted on the measured angular power spectra from DESI LRG sample in section~\ref{sec:results_LRG}.

For each input data, we run the algorithm with 1000 initial matrices, but only 100 matrices with lowest $\mathcal{J}$ will be input to Algorithm 2 from the output of Algorithm 1 (see section~\ref{sec:nmf}).
Then, we select the results via
$\mathcal{J}/\mathcal{J}_{\min}-1\leq 10~\%$.
The $\mathcal{J}_{\min}$ is the minimum objective function among the final output of algorithm from 100 initial matrices.
Using the solutions with objective function in this range, we take the median values of scattering matrix elements as the final result, with renormalization to 1 in each column.
We note that here the normalized median value of each scattering matrix element typically matches the mean value, provided that the reconstructions are successful.
However, the performance of the normalized median absolute deviation $\sigma_{\mathrm{NMAD}} \approx 1.48 \times \operatorname{median}(|P^{\rm recon}_{ij}-\operatorname{median}(P^{\rm recon}_{ij})|)$, may distinctly differ from that of the standard deviation due to a few outliers.
We adopt the normalized median value and $\sigma_{\mathrm{NMAD}}$ of each scattering matrix element as the fiducial results in this paper.
We concentrate on redshift distributions, but the auto power spectra in true-$z$ bins can also be accurately derived simultaneously.

\subsection{Results from mock datasets}\label{sec:results_mock}
\begin{figure}
\centering
    \begin{minipage}{7.5cm}
        \centering
        \includegraphics[width=7.5cm]{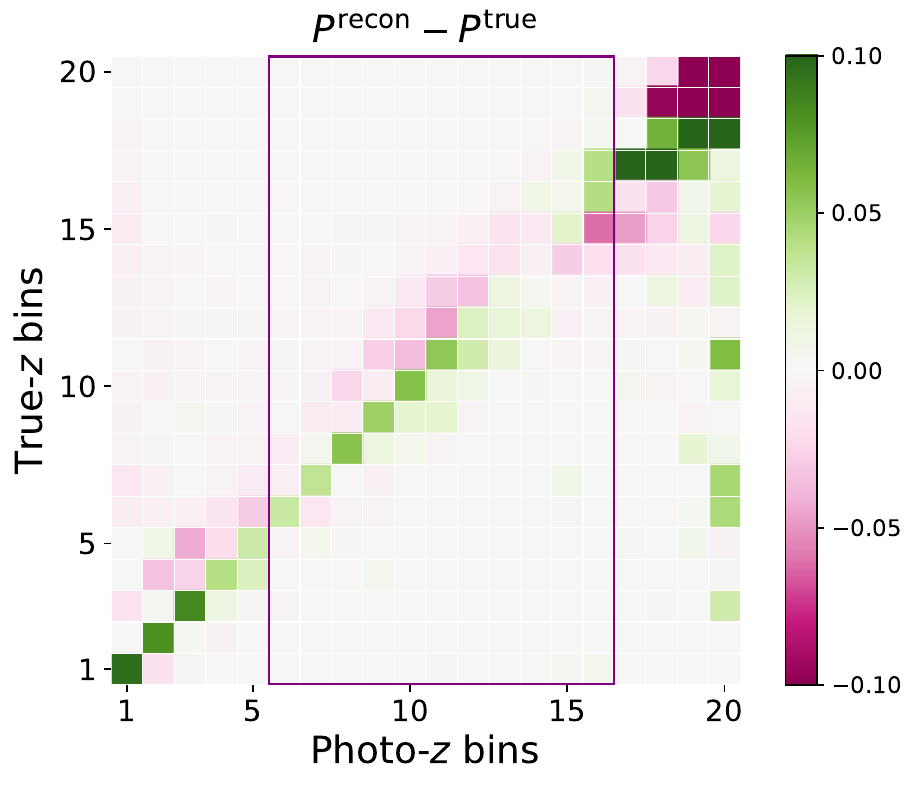}
    \end{minipage}
    \hspace{0.2cm}
	\begin{minipage}{7.5cm}
    	\centering
        \includegraphics[width=7.5cm]{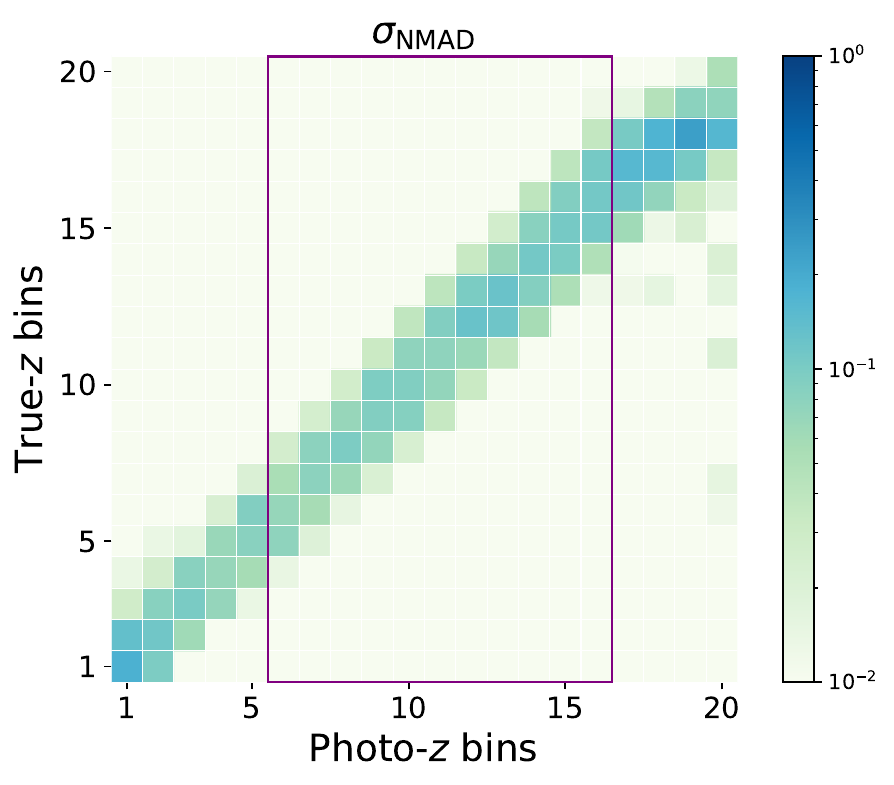}
    \end{minipage}
    \begin{minipage}{7.5cm}
        \centering
        \includegraphics[width=7.5cm]{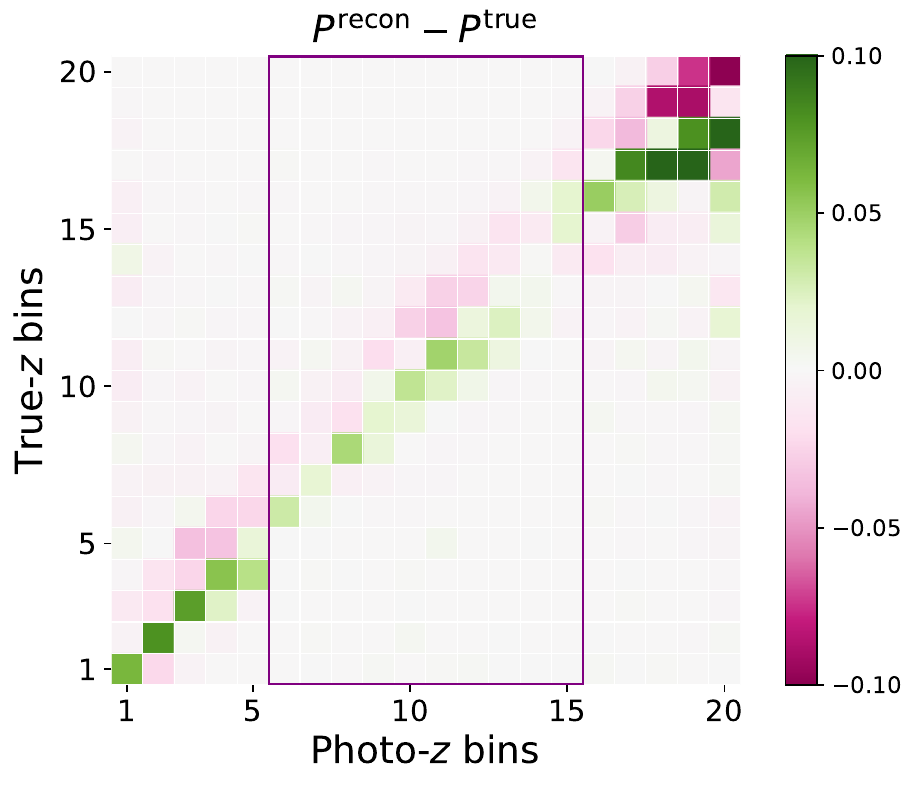}
    \end{minipage}
    \hspace{0.2cm}
	\begin{minipage}{7.5cm}
    	\centering
        \includegraphics[width=7.5cm]{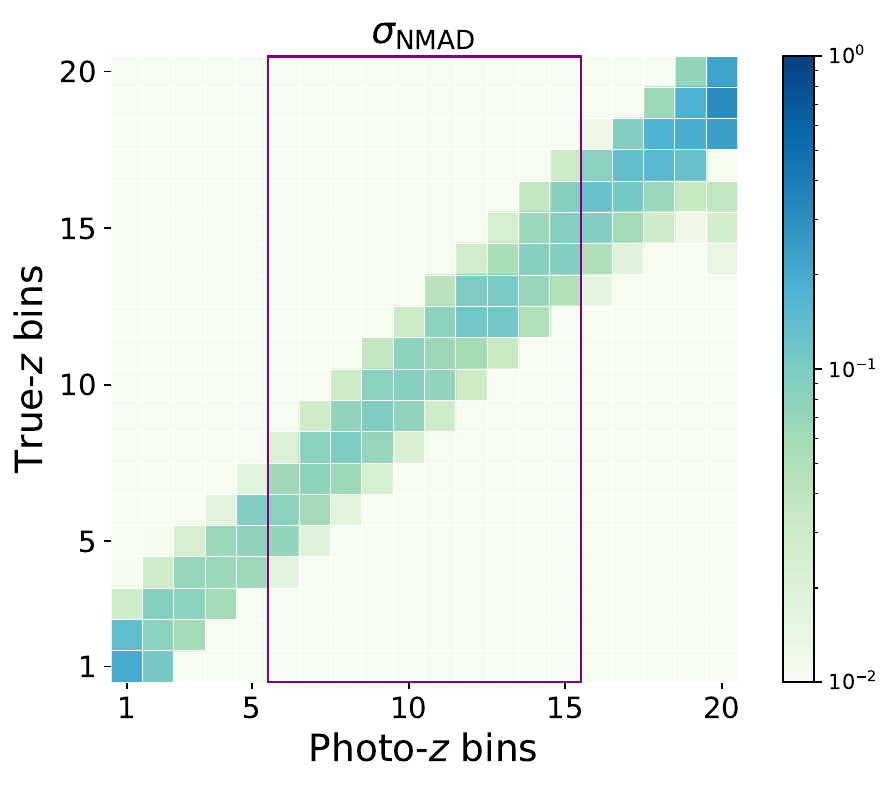}
    \end{minipage}
    \caption{The reconstruction results from mock data characterized by Low Photo-$z$ Error with each redshift bin width set to 0.06, are presented for High Noise (upper) and Low Noise (lower).
    The left and right indicate the biases and $1\sigma$ uncertainties of the reconstructed scattering matrices, respectively.
    For each panel, the vertical axis labels the true redshift bins, the horizontal axis denotes the photometric redshift bins.
    The region enclosed by the purple line corresponds to the photo-$z$ bins selected using a  criterion based on the relative values of the uncertainties, which is presented in section~\ref{sec:results_mock}.
    }
    \label{fig:mock_p_error0.03_dz0.06}
\end{figure}

\begin{figure}
\centering
    \begin{minipage}{7.5cm}
        \centering
        \includegraphics[width=7.5cm]{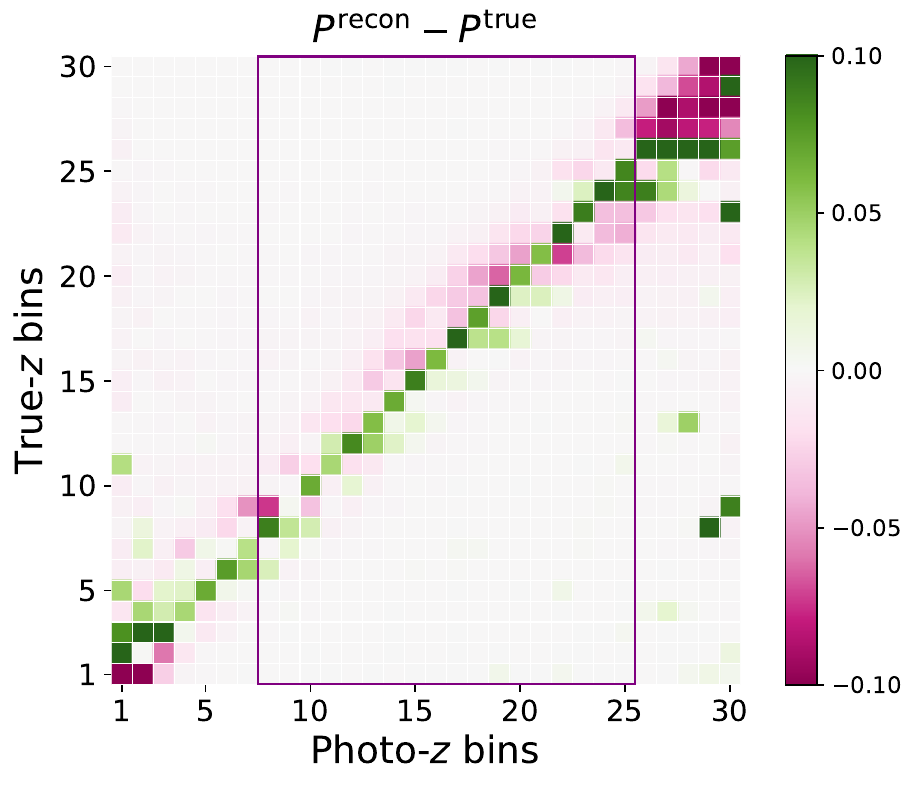}
    \end{minipage}
    \hspace{0.2cm}
	\begin{minipage}{7.5cm}
    	\centering
        \includegraphics[width=7.5cm]{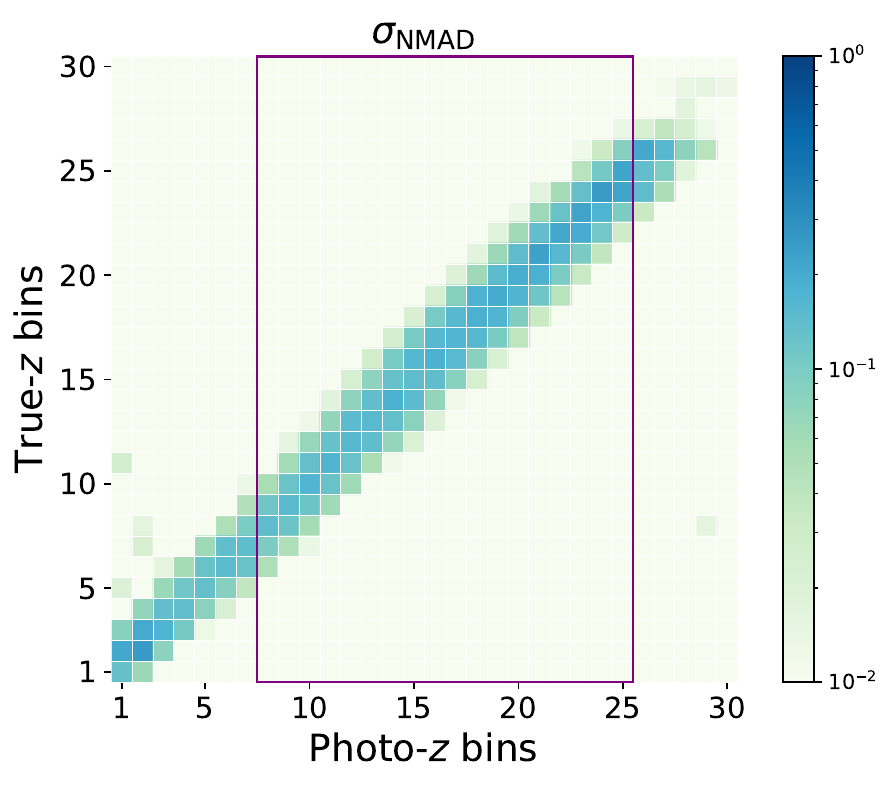}
    \end{minipage}
    \begin{minipage}{7.5cm}
        \centering
        \includegraphics[width=7.5cm]{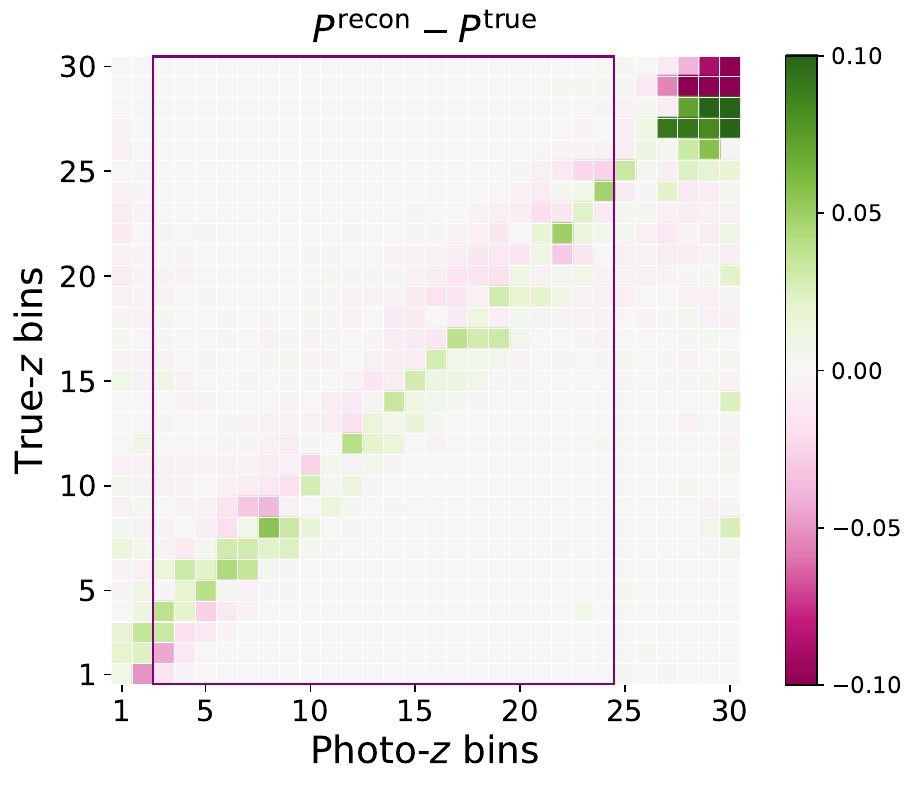}
    \end{minipage}
    \hspace{0.2cm}
	\begin{minipage}{7.5cm}
    	\centering
        \includegraphics[width=7.5cm]{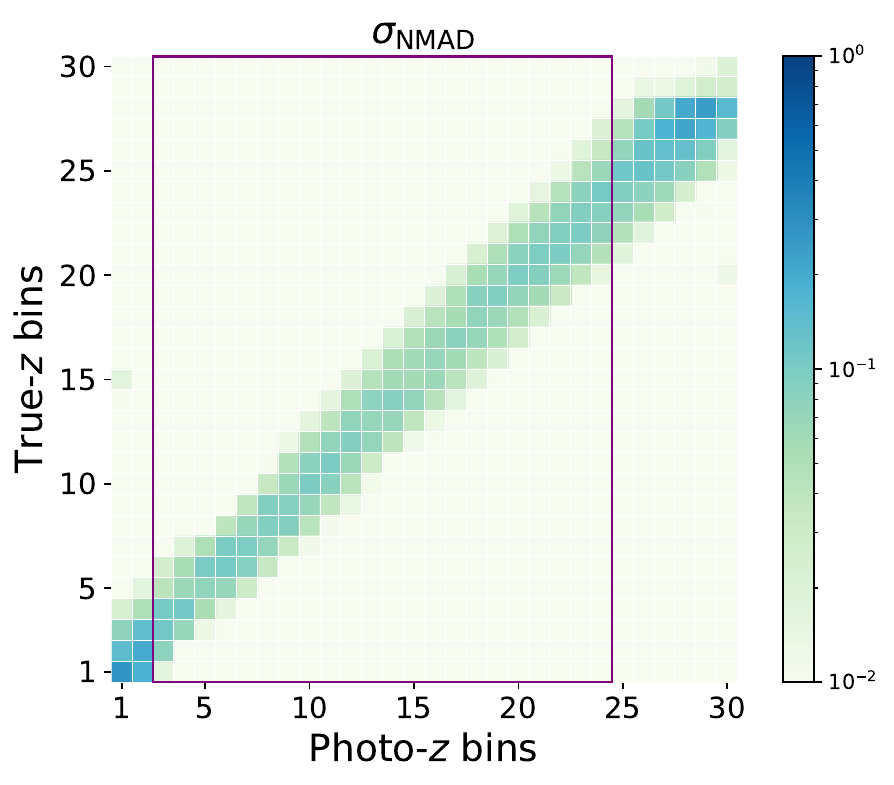}
    \end{minipage}
    \caption{Same as figure~\ref{fig:mock_p_error0.03_dz0.06}, but the redshift bin widths are set to 0.04.}
    \label{fig:mock_p_error0.03_dz0.04}
\end{figure}

\begin{figure}
\centering
    \begin{minipage}{7.5cm}
        \centering
        \includegraphics[width=7.5cm]{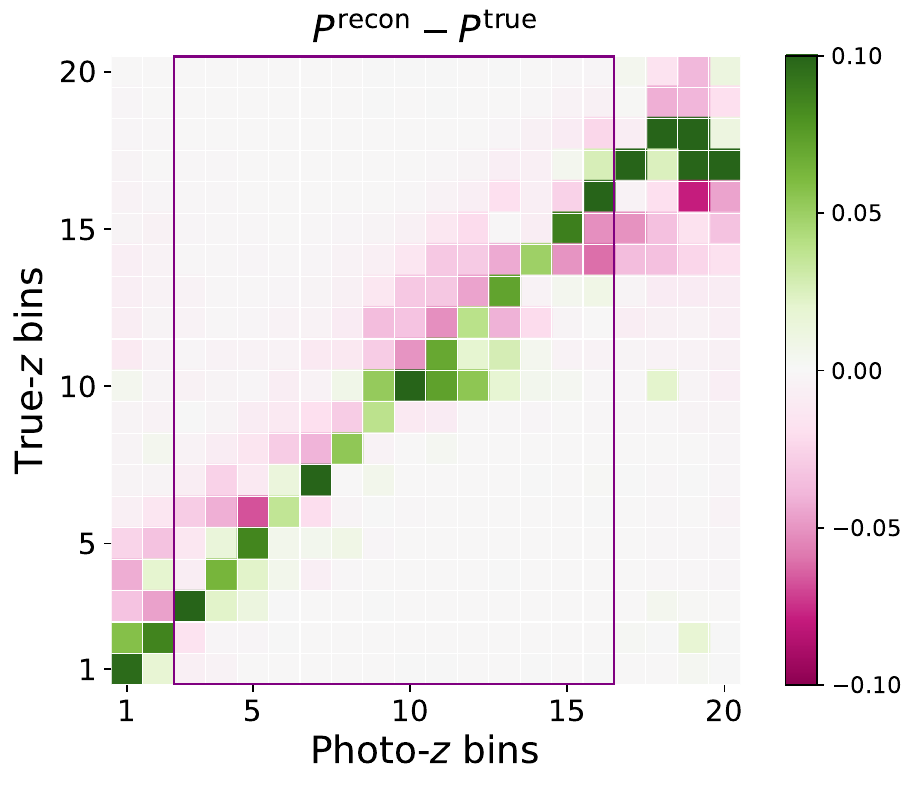}
    \end{minipage}
    \hspace{0.2cm}
	\begin{minipage}{7.5cm}
    	\centering
        \includegraphics[width=7.5cm]{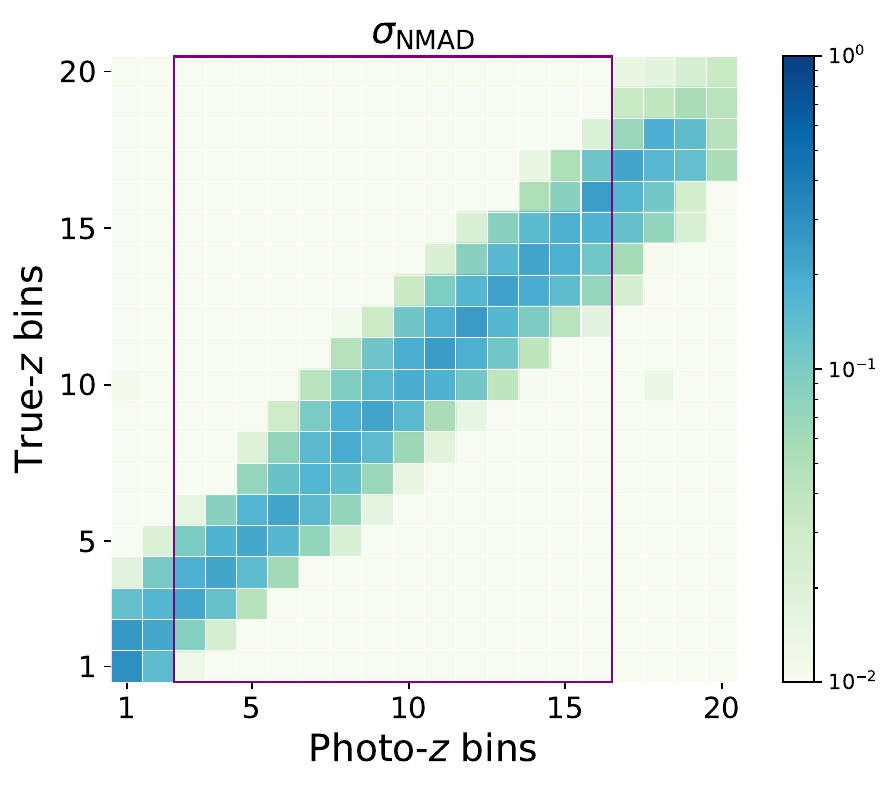}
    \end{minipage}
    \begin{minipage}{7.5cm}
        \centering
        \includegraphics[width=7.5cm]{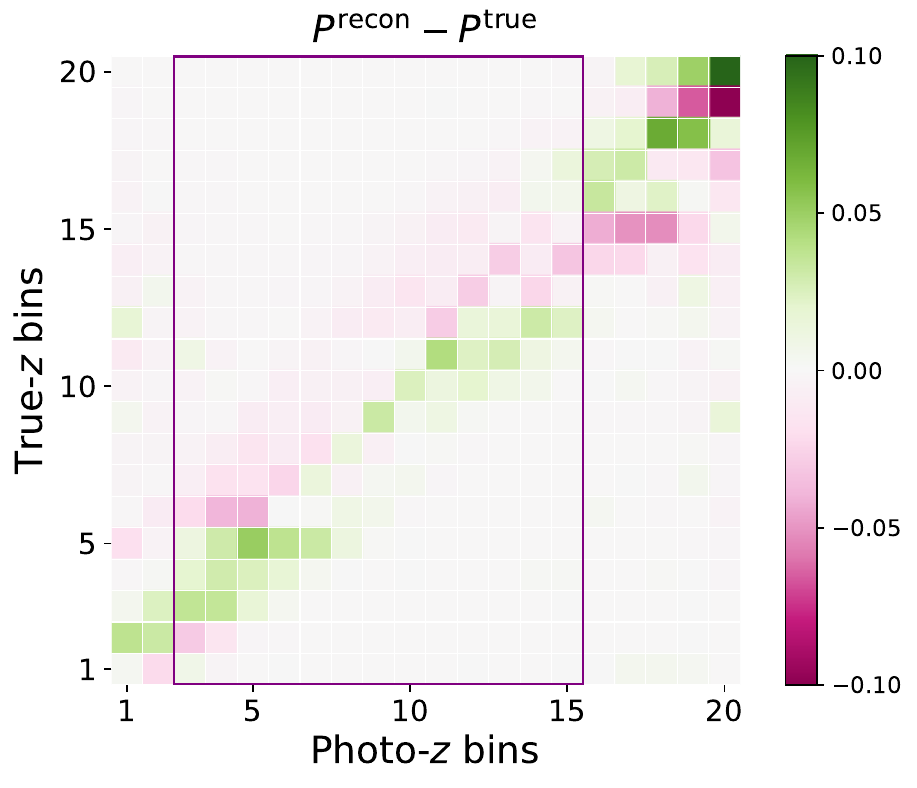}
    \end{minipage}
    \hspace{0.2cm}
	\begin{minipage}{7.5cm}
    	\centering
        \includegraphics[width=7.5cm]{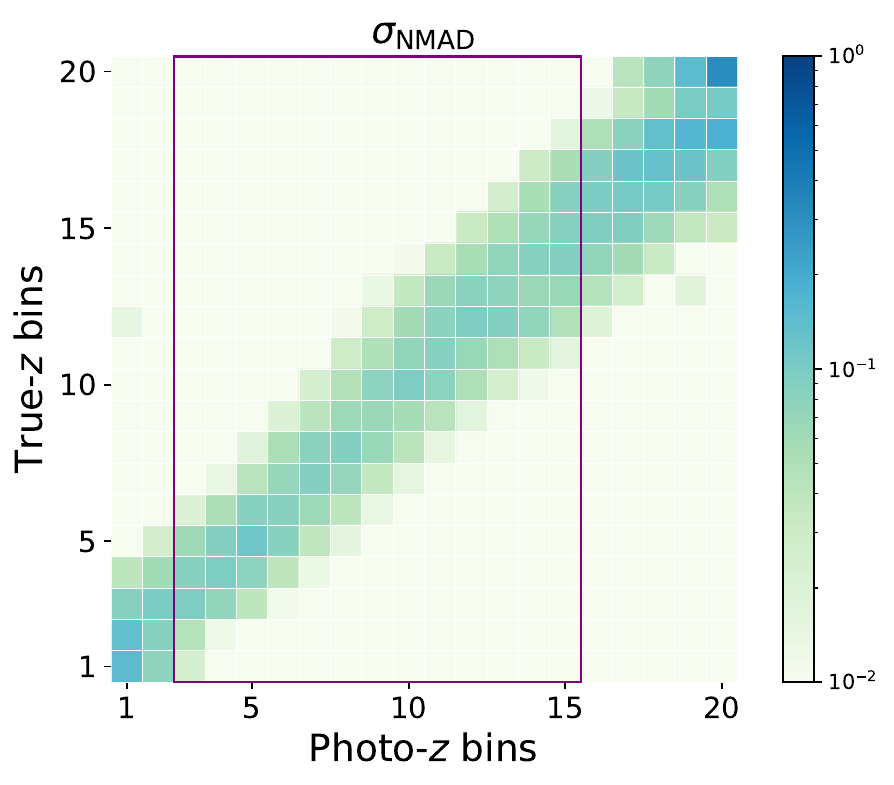}
    \end{minipage}
    \caption{Same as figure~\ref{fig:mock_p_error0.03_dz0.06}, but for High Photo-$z$ error.}
    \label{fig:mock_p_error0.05_dz0.06}
\end{figure}

\begin{table*}
	\centering
	\caption{Summary of the mean redshift biases in the photometric redshift bins for different mock datasets. The five equal-width sequential photometric redshift bins have widths set to 0.12, and the total redshift ranges for the cases in figure~\ref{fig:mock_p_error0.03_dz0.06} and figure~\ref{fig:mock_p_error0.03_dz0.04} are [0.40, 1.00] and [0.42, 1.02], respectively.}
	\label{tab:mean_z_result}
    \vspace{10pt}
    \scriptsize
	\resizebox{13cm}{!}{
	\begin{tabular}{ccccccc}
		\hline		  \multicolumn{2}{c}{\multirow{2}{*}{Case}}&\multicolumn{5}{c}{Mean redshift bias $|\Delta z|/(1+z)$}\\
		\cline{3-7}
		  &&Bin 1& Bin 2 & Bin 3 & Bin 4 & Bin 5\\
		\hline
		\multirow{2}{*}{Figure~\ref{fig:mock_p_error0.03_dz0.06}} &High Noise&$0.004$&$0.005$&$0.007$ &$0.004$ &$0.0007$ \\
        &Low Noise&$0.003$ &$0.003$ &$0.005$  &$0.004$ &$0.0005$ \\
		\hline
		\multirow{2}{*}{Figure~\ref{fig:mock_p_error0.03_dz0.04}}&High Noise& $0.005$ &$0.007$ &$0.008$ &$0.007$ &$0.002$ \\
        &Low Noise&$0.004$ &$0.004$ &$0.004$ &$0.003$ &$0.001$ \\
		\hline
	\end{tabular}}
\end{table*}
Despite conducting a range of data tests that varied in photo-$z$ errors, noise level, and redshift bin width, only a subset of these results are showcased in this paper.
Initially, we apply the algorithm to the mock data characterized by Low Photo-$z$ Error and both two noise level described in section~\ref{sec:mock_data}.
The redshift range from 0.1 to 1.3 is divided into 20 bins, each with a width of 0.06.
The results of biases on scattering matrices ($P^{\rm recon}-P^{\rm true}$) and $\sigma_{\mathrm{NMAD}}$ are shown in figure~\ref{fig:mock_p_error0.03_dz0.06}.
We find that accurate reconstructions can be achieved by using the self-calibration algorithm with its newly integrated update rules.
The mean absolute bias across all elements of the reconstructed scattering matrix is approximately 0.011 (0.008) for the case with High Noise (Low Noise).
The comparison results demonstrate that the reconstruction accuracy can be obviously improved by expanding the area and increasing the number density.
Consequently, we anticipate that the reconstruction results for stage-IV cosmic shear samples will exhibit enhanced accuracy.

We note that cosmological analyses commonly concentrate on redshift regions with high galaxy number density, and the relatively large reconstruction errors at both ends of the redshift range are not expected to have a considerable impact.
To streamline the application of reconstruction results, we also introduce a concise yet rigorous selection criterion here (depicted in the purple frame of figure~\ref{fig:mock_p_error0.03_dz0.06}).
This criterion is established through a three-step process:
\vspace{1em}

(i) the median half of all the photo-$z$ bins are selected to represent the most conservative outcome.

(ii) we determine the mean value ($\mu_c$) and standard deviation ($\sigma_c$) of $\sigma_{\mathrm{NMAD}}$ from the sum of the diagonal and sub-diagonal elements for each photo-$z$ bin within this redshift range.

(iii) we extend the region from both sides, ensuring that the average value of the sum of the diagonal and sub-diagonal elements from the newly added two photo-$z$ bins falls within the range $[\mu_c - \sigma_c,\,\mu_c + 3\sigma_c]$.
Due to the typically large reconstruction uncertainties in low-density redshift bins, particularly in regions close to the redshift edges, the algorithm has a relatively high probability of finding a local minimum solution, resulting in a rapid decrease in $\mu_c$.
Thus, we set this stringent lower limit to prevent issues that could arise from any potential local minima.
\vspace{1em}

Additionally, we directly exclude two photo-$z$ bins closest to each redshift edge for two primary reasons: (i) the unacceptable noise stemming from low density; (ii) the inevitable presence of sources outside the redshift range in realistic scenarios, which can markedly degrade the reconstruction accuracy on these photo-$z$ bins.
Within this selected region, the mean absolute bias of elements is reduced to $\sim$0.006 ($\sim$0.004) for High Noise (Low Noise).
It is important to note that the criterion we present is predicated on the relative values of the uncertainties.
When applying this criterion to realistic cases, adjustments can be made by taking into account their absolute magnitudes, allowing for a more flexible or stringent approach as needed.

In order to demonstrate the reconstruction performance more adequately and to compare with the requirements of surveys, we consider five equal-width sequential test photometric redshift bins within $0.4\leq z_p<1.0$.
The widths of these test redshift bins are set to 0.12 to match the practical data analysis in surveys, where each bin width typically exceeds 0.1.
We note that this redshift range is within the selected region.
Then we combine the corresponding photo-$z$ bins from the current thin binning scheme of the reconstructed data, and obtain the redshift distributions of five broad photo-$z$ bins.
We present the mean redshift biases for these five photo-$z$ bins in table~\ref{tab:mean_z_result}, and it is obvious that the mean redshift of each bin can be accurately reconstructed with the self-calibration algorithm.
Although the reconstruction accuracy may fall short of the requirements for the stage-IV surveys (e.g., $0.002(1+z)$ for LSST Year 1 weak lensing analysis), it is comprehensible given the much higher noise levels.

Considering the self-calibration method is more robust and applicable when applied to thin redshift bins.
In figure~\ref{fig:mock_p_error0.03_dz0.04} we present the results with the redshift bin widths adjusted to 0.04, while keeping all other aspects of the analysis unchanged.
The mean absolute bias across all elements of the reconstructed scattering matrix stands at approximately 0.011 (0.006) for High Noise (Low Noise), which is further reduced to around 0.008 (0.004) within the selected region.
Furthermore, the mean redshift biases for five equal-width test photometric redshift bins (within $0.42\leq z_p<1.02$) are also presented in table~\ref{tab:mean_z_result}.
Here, the total redshift range varies slightly due to the binning scheme adopted.
We find that the impact of noise is significantly greater here, which could potentially lead to failures in reconstruction within a thinner binned scheme.
When implementing the self-calibration algorithm, we can minimize the redshift bin size to the greatest extent possible until the reconstruction result become messy (instead of distributed around the diagonal region) or the algorithm encounters a crash.
The accuracy can also be self-checked by combing the reconstructed results from different binning schemes to some identical tomographic bins.

Figure~\ref{fig:mock_p_error0.05_dz0.06} demonstrates the reconstructions on the cases with High Photo-$z$ Error.
In appendix~\ref{sec:appendix_mock_results} we investigate the ability of the algorithm in the context of a scenario involving a catastrophic overall offset in photo-$z$ error.
The still stable performance of results shows the robust capability of the self-calibration algorithm to handle a variety of complex conditions.

\subsection{Application to DESI LRG sample}\label{sec:results_LRG}
\begin{figure}
\centering
    \begin{minipage}{7.5cm}
        \centering
        \includegraphics[width=7.5cm]{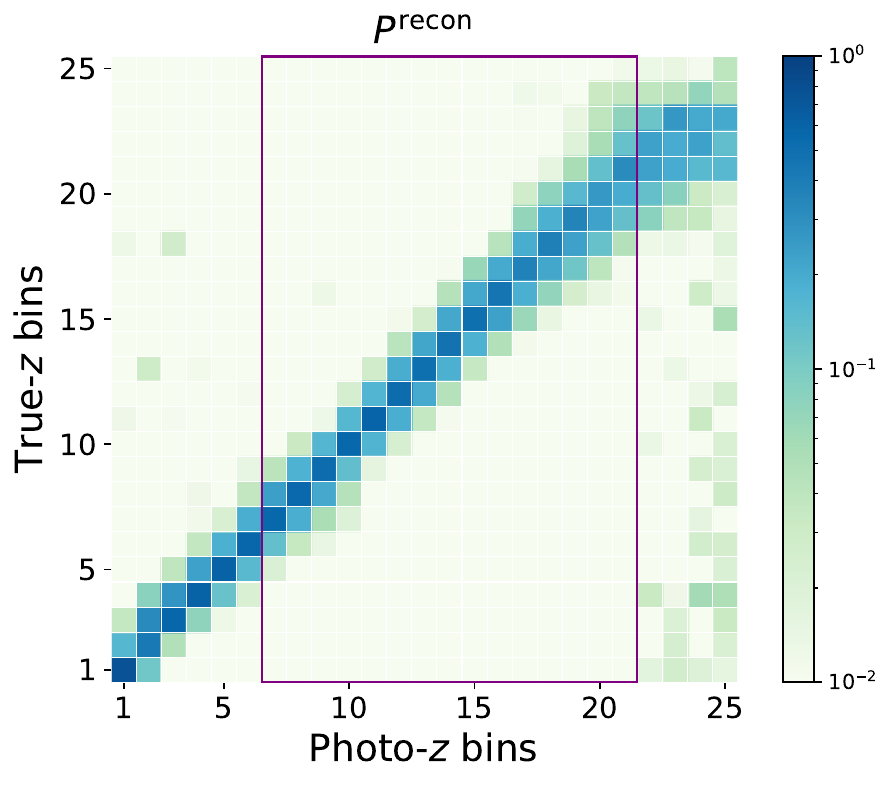}
    \end{minipage}
    \hspace{0.2cm}
	\begin{minipage}{7.5cm}
    	\centering
        \includegraphics[width=7.5cm]{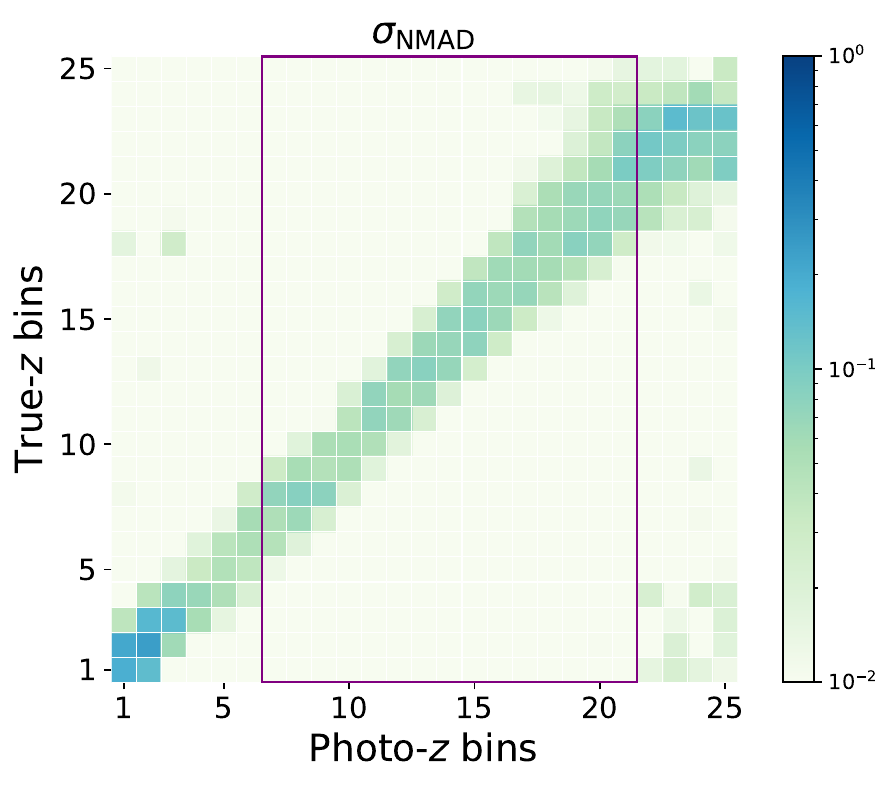}
    \end{minipage}
    \caption{
    Left: The scattering matrix reconstructed from the self-calibration algorithm when applying to DESI LRG sample.
    Right: The $1\sigma$ uncertainties determined by the selected reconstruction results from different perturbated photometric galaxy-galaxy power spectra.
    The majority of bin widths are set to 0.04, except for the two broad bins on either end and those that are specifically designed to retain the boundaries of subsamples.
    The photo-$z$ bins enclosed by the purple line are selected using the same criterion applied to the mock data.
    }
    \label{fig:LRGs_recon_dz0.04}
\end{figure}
\begin{figure}
\centering
    \begin{minipage}{4.5cm}
        \centering
        \includegraphics[width=4.3cm]{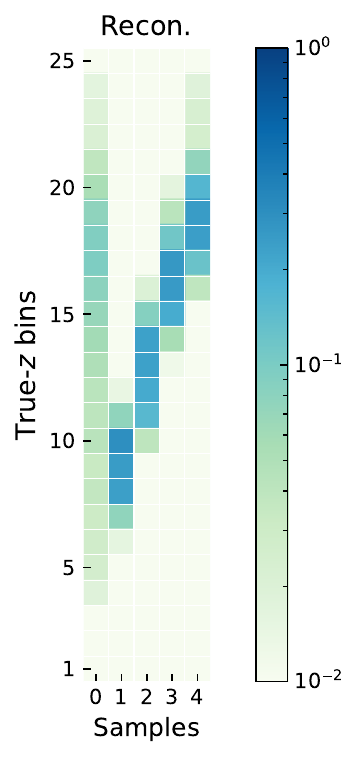}
    \end{minipage}
    \hspace{0.1cm}
	\begin{minipage}{4.5cm}
    	\centering
        \includegraphics[width=4.3cm]{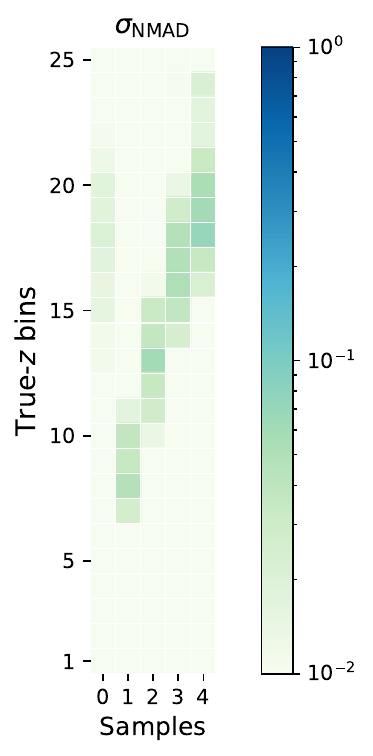}
    \end{minipage}
    \hspace{0.1cm}
	\begin{minipage}{4.5cm}
    	\centering
        \includegraphics[width=4.5cm]{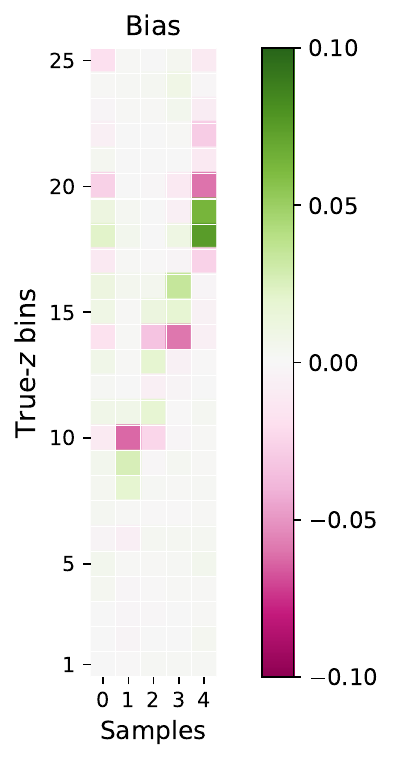}
    \end{minipage}
    \caption{The reconstructions and comparison with the benchmark references of five samples (full DESI LRG sample and four subsamples).
    The left, middle, and right panels respectively illustrate the reconstructed distributions, the associated uncertainties, and the biases in comparison to the benchmark references.
    }
    \label{fig:LRG_smaples_recon}
\end{figure}
\begin{figure}
    \centering
	\includegraphics[width=13cm]{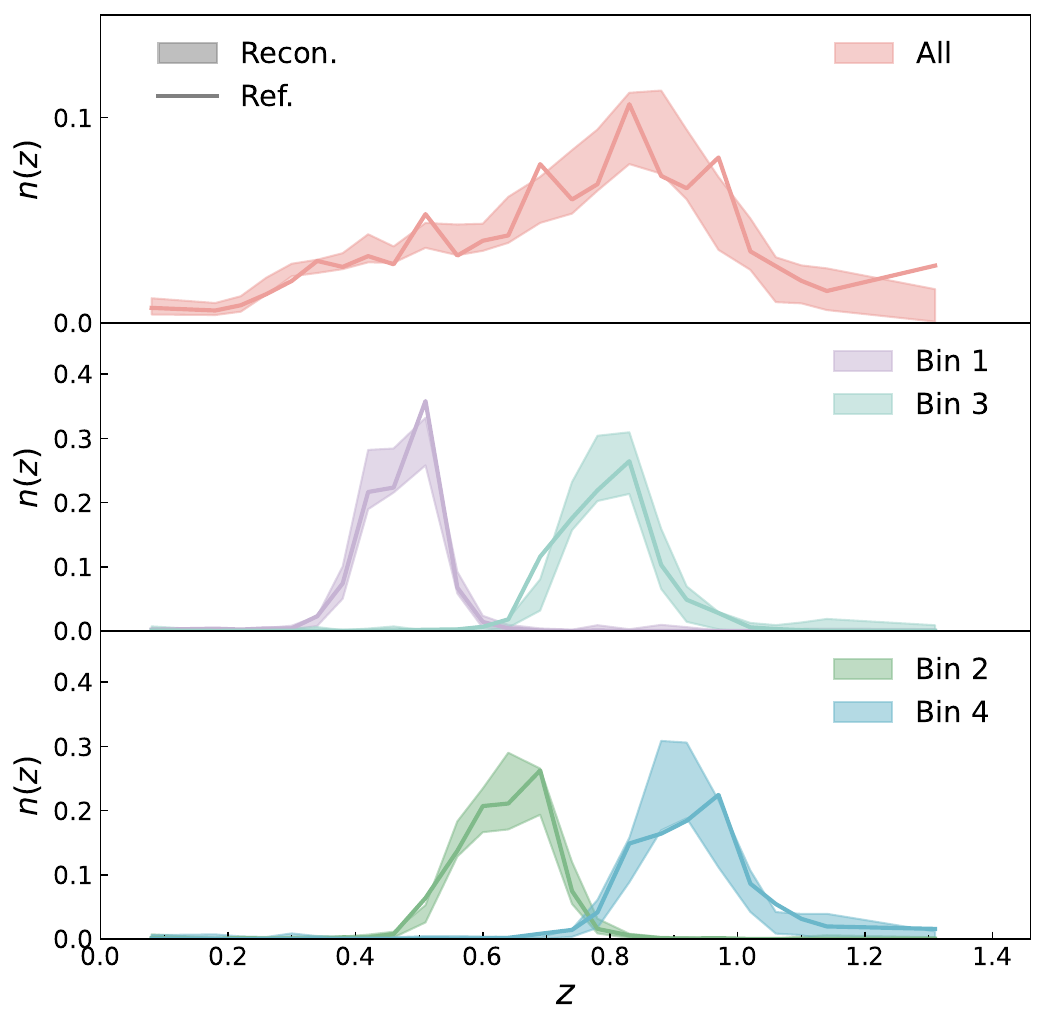}
    \caption{
    Visual representation of redshift distributions reconstructed from self-calibration algorithm using photometric galaxy clustering (lighter shades) and the benchmark references using DESI SV1 spectroscopic redshifts (solid lines) in \cite{Zhou:2023aa}, for full DESI LRG sample and four subsamples.
    The differences in distribution shape and range compared to figure~\ref{fig:n_z} are a result of the binning scheme that was adopted, including two broad bins on either side and those specifically designed to preserve the boundaries of subsamples.
    The bands indicate the 1$\sigma$ uncertainties from the central value.
    }
    \label{fig:LRG_smaples_nz_dz0.04}
\end{figure}

Following the successful implementation and testing across a range of mock datasets, we proceed to apply the algorithm to the DESI LRG sample.
To facilitate a comparison with the referenced redshift distributions in \cite{Zhou:2023aa}, we maintain the boundaries of four subsamples in our binning scheme, thereby dividing the whole redshift range ($0<z<1.46$) into 25 tomographic bins with edges $z_p=$ 0, 0.16, 0.20, 0.24, 0.28, 0.32, 0.36, 0.40, 0.44, 0.48, 0.54, 0.58, 0.62, 0.66, 0.72, 0.76, 0.80, 0.86, 0.90, 0.94, 1.00, 1.04, 1.08, 1.12, 1.16, and 1.46.
The majority of bin widths are established at 0.04, with the exception of the two broad bins on either side and those specifically designed to preserve subsamples ($z_p=$ 0.40, 0.54, 0.72, 0.86, and 1.0 as bin edges).
The reconstructed results are shown in figure~\ref{fig:LRGs_recon_dz0.04}, alongside the photo-$z$ bins selected using the same criterion applied to the mock data.
We find that the redshift ranges of the four subsmples fall within our selected region, which suggests a high level of reconstruction accuracy.
The uncertainty here is obviously lower than the similar case shown in the upper right panel of figure~\ref{fig:mock_p_error0.03_dz0.04}, and the direct connection on accuracy is completely inadvisable due to the differences in redshift error form, galaxy bias, binning scheme, and other factors.

We denote the full DESI LRG sample as Sample 0 and the four subsamples as Samples 1-4, and the correspondences to the current binning scheme in figure~\ref{fig:LRGs_recon_dz0.04} are:
\vspace{1em}
\begin{itemize}
    \item Sample 0 (All, $0<z_p<1.46$): bins [1, 25].
    \item Sample 1 (Bin 1, $0.40\leq z_p<0.54$): bins [8, 11).
    \item Sample 2 (Bin 2, $0.54\leq z_p<0.72$): bins [11, 15).
    \item Sample 3 (Bin 3, $0.72\leq z_p<0.86$): bins [15, 18).
    \item Sample 4 (Bin 4, $0.86\leq z_p<1.00$): bins [18, 21).
\end{itemize}
\vspace{1em}
Reconstructions on the photo-$z$ bins near the redshift edges might be less accurate, but this is unlikely to significantly impact the redshift distribution of the full DESI LRG sample due to the small number of galaxies within those bins.
By combining corresponding photo-$z$ bins from current binning scheme, figure~\ref{fig:LRG_smaples_recon} presents the reconstructed results of five samples and the biases in relation to the benchmark references in \cite{Zhou:2023aa}.
The mean absolute bias from the references on the reconstructed results in figure~\ref{fig:LRG_smaples_recon} is about 0.008.
As previously noted, the $P_{ij}$ scatter carries the same implications as the redshift distribution.
To better visualize our results, we show the comparisons in the view of redshift distributions in figure~\ref{fig:LRG_smaples_nz_dz0.04}.
It is obvious that the self-calibration algorithm can reconstruct accurate redshift distributions using only photometric galaxy clustering, which are comparable to the benchmark references using spectroscopic redshifts from the DESI SV1 data.
The biases of mean redshifts for full DESI LRG sample and four subsamples between the reconstructions and references are $\sim0.012(1+z),~\sim0.005(1+z),~\sim0.002(1+z),~\sim0.003(1+z),~\sim0.014(1+z)$, respectively.
Note that there are inevitable uncertainties in the references due to the inherent limitations of the spectroscopic redshift data.
We anticipate that combining the self-calibration method with other statistics, as detailed in the subsequent section, will significantly enhance accuracy and reduce uncertainty.
In order to illustrate the generality and robustness, in appendix~\ref{sec:appendix_LRG_results} we also present the reconstructed redshift distributions using another redshift binning scheme with the majority of bin widths set to 0.06.

A noteworthy consideration is that the analysis in this paper does not account for potential systematics in the galaxy clustering signal, such as lensing magnification.
The results obtained suggest that the impact of these systematic errors is not significant, at least for the DESI LRG sample under consideration.
Additionally, the 2$\sigma$ cut utilized in DES analysis \cite{Gatti:2018tg,Gatti:2022va} for excluding the tails of the redshift distributions can be applied to reduce the impact of magnification on mean redshift estimation.
Although adequate for method validation, future applications for more nuanced analysis might necessitate additional operations to minimize the possible effects.

\section{Discussion and conclusions}\label{sec:dis_cons}
We presented an improved self-calibration algorithm, equipped with novel update rules, capable of accurately reconstructing redshift distributions in tomographic bins through photometric galaxy clustering.
Incorporating the ability to handle nonuniform uncertainties and negative data elements introduced by noise leads to notable improvements in both the accuracy and applicability of the self-calibration method.
We emphasized the necessity of the reconstruction on thin tomographic bins to prevent theoretical failure.
We found that the algorithm succeeds in robust and accurate performance on various mock data and the reconstructed redshift distributions of DESI LRG sample are comparable to the state-of-the-art ones using DESI SV1 spectroscopic redshifts.
The results indicate that the self-calibration method holds great promise for cosmological analyses involving various photometric survey data, particularly well-suited for addressing scenarios where an appropriate spectroscopic sample is lacking for calibration purposes.

The self-calibration algorithm is highly user-friendly for application, as it requires only the galaxy-galaxy correlations between photometric redshift bins as input, eliminating the need for any complex or preliminary procedures.
Furthermore, the self-calibration technique based on galaxy clustering can also handle other problems besides photo-$z$ error, such as the interloper bias in spectroscopic surveys \cite{Peng:2023aa}.
Consequently, the novel update rules we propose can also be integrated to improve the accuracy of the reconstruction there.
The algorithm, featuring modifications, is renewed for public use at \href{https://github.com/PengHui-hub/FP-NMF}{https://github.com/PengHui-hub/FP-NMF}.

While these results are encouraging, considering the limiting factors of our method can provide valuable insights for future work.
As previously emphasized, it is essential to minimize the redshift bin size, unless doing so leads to a significant deterioration in the reconstruction results or causes the algorithm to crash.
The constraints derived from photometric galaxy clustering are inevitably sensitive to the number density within each tomographic bin.
Although cosmological analyses typically focus on redshift regions with high number density, caution is advised when establishing the binning scheme.
This is to prevent a potential domino effect stemming from the influence of low-density redshift bins on either side.
For typical samples in galaxy surveys, employing two broad bins may offer a viable way.
Moreover, the photometric redshift quality should not be so poor that it causes substantial sub-dominant diagonal elements in scattering matrix, as this would lead to remarkable degradation in reconstruction.
The selection criterion introduced in section~\ref{sec:results_mock} allows for flexible adjustment to suit the requirements of the analysis and to align with the overall magnitude of uncertainty.

This self-calibration method serves as a significant supplement to the current approaches for redshift calibration.
In order to meet the challenges arising from the requirements of precise cosmology, we outline two main opportunities for further enhancing the accuracy as follows:
\vspace{1em}

(i) spectroscopic redshift sample: Since the self-calibration technique is entirely independent of spectroscopic data, it possesses considerable potential for combination with other existing methods, particularly the cross-correlation with reference galaxy samples.

(ii) weak lensing data: Lensing-galaxy correlations are highly beneficial for breaking certain degeneracies from galaxy-galaxy correlations and ultimately enhancing the accuracy of the self-calibration method \cite{Zhang:2010wr}.
\vspace{1em}

Given the reliable performance of the current self-calibration method, any successful combination with other data sources would be quite meaningful.
We intend to conduct these challenging but rewarding studies in future work.

\appendix
\section{Comparison with previous algorithm}\label{sec:appendix_compare_algorithm}
\begin{figure*}
\centering
    \begin{minipage}{7.5cm}
        \centering
        \includegraphics[width=7.5cm]{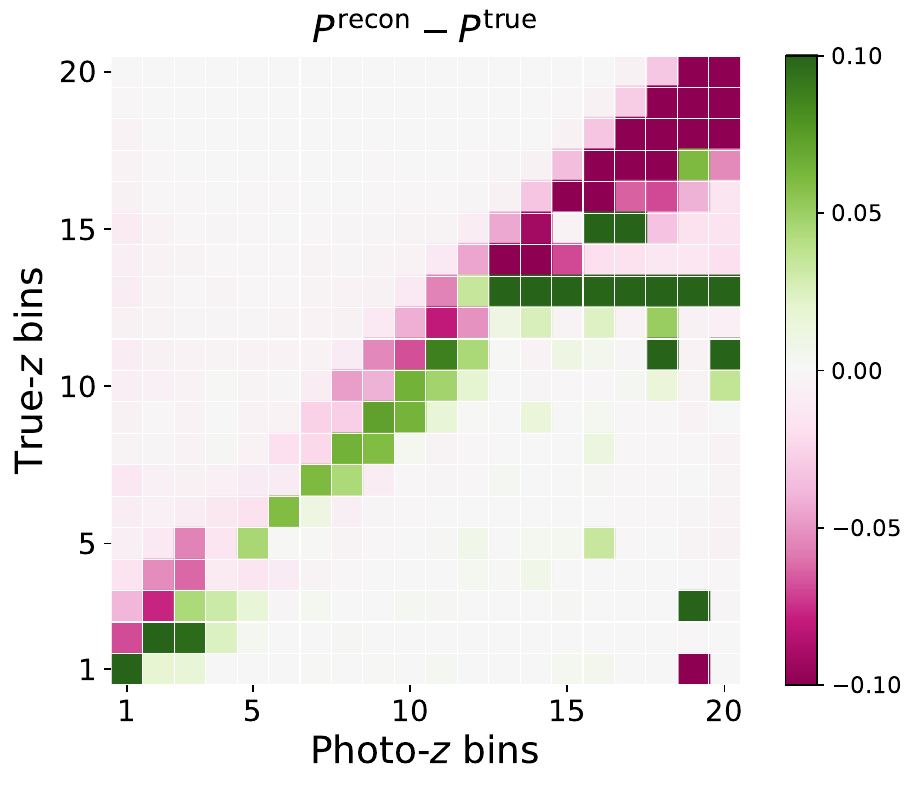}
    \end{minipage}
    \hspace{0.2cm}
	\begin{minipage}{7.5cm}
    	\centering
        \includegraphics[width=7.5cm]{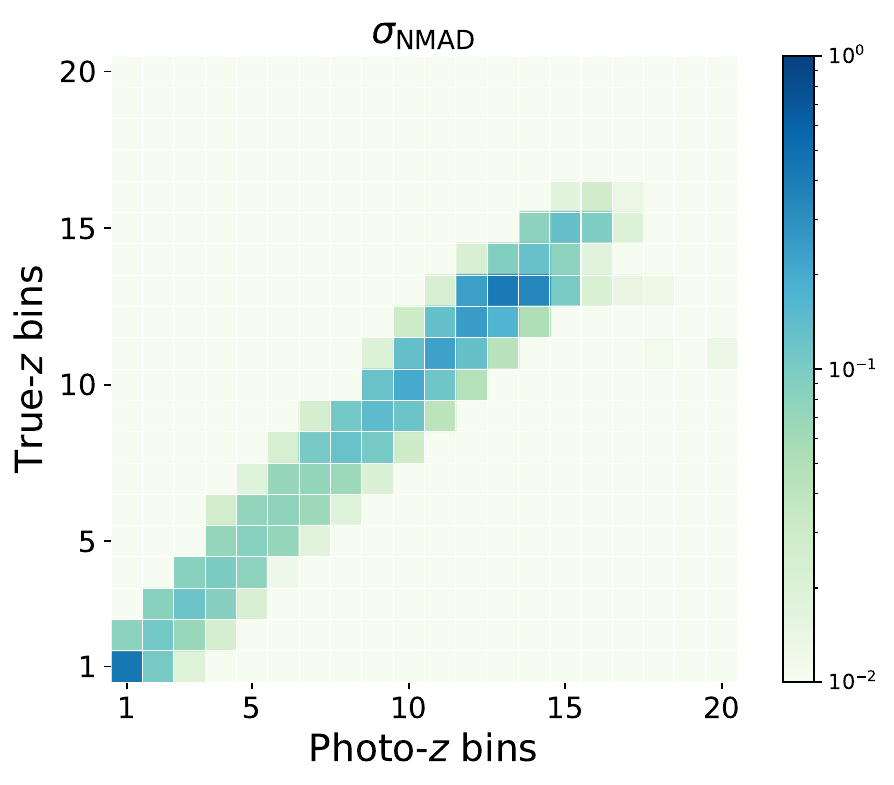}
    \end{minipage}
    \caption{
    Reconstructions when applying previous self-calibration algorithm to the mock data, which is same as the case of upper panels in figure~\ref{fig:mock_p_error0.03_dz0.06}.
    The left and right panel indicate the bias and $1\sigma$ uncertainty of the reconstructed scattering matrix, respectively.
    }
    \label{fig:compare}
\end{figure*}
Here, we compare the reconstructions using previous algorithm on mock data same to the case of upper panels in figure~\ref{fig:mock_p_error0.03_dz0.06}.
We note that all the operation are the same, except the update rules and objective function in the algorithm.
The results shown in figure~\ref{fig:compare} suggest that the previous algorithm struggles to satisfy the demands of realistic applications.
The improvements we have introduced are not only urgent but also crucial for broadening and advancing the application of the self-calibration method.

\section{Mock data with an overall offset redshift error}\label{sec:appendix_mock_results}
\begin{figure*}
\centering
    \begin{minipage}{7.5cm}
        \centering
        \includegraphics[width=7.5cm]{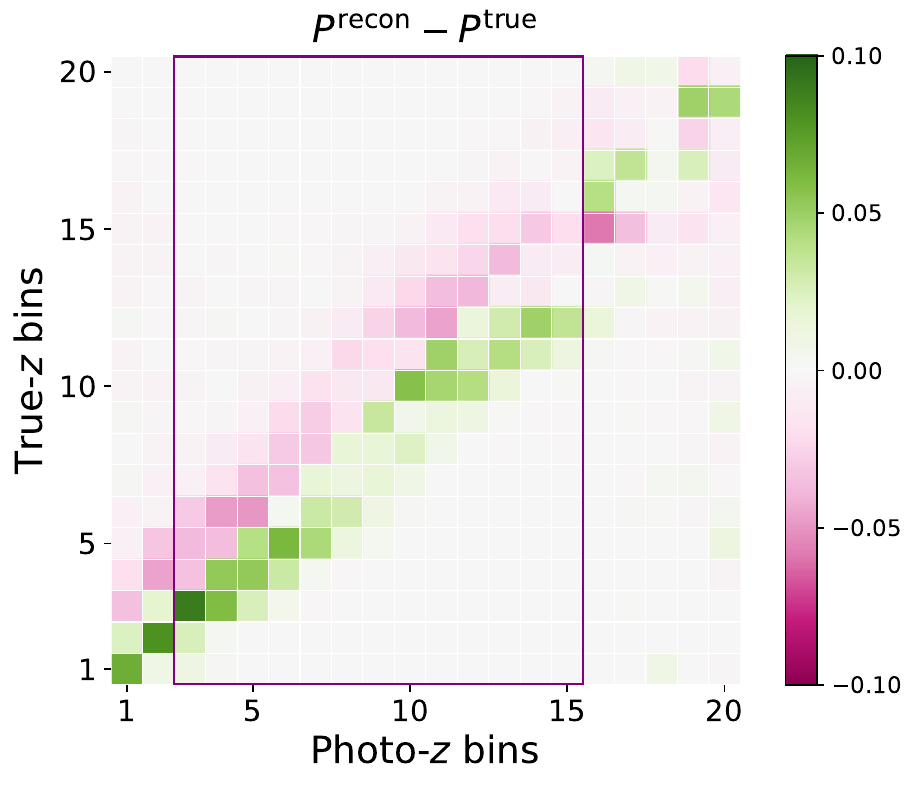}
    \end{minipage}
    \hspace{0.2cm}
	\begin{minipage}{7.5cm}
    	\centering
        \includegraphics[width=7.5cm]{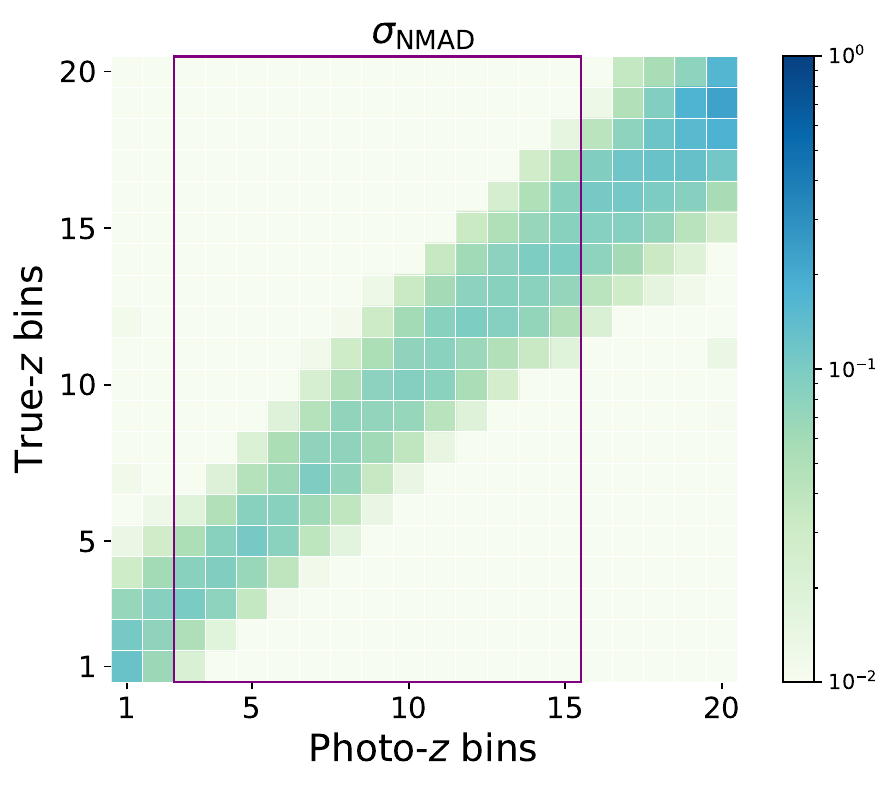}
    \end{minipage}
    \caption{
    Same as the lower panels in figure~\ref{fig:mock_p_error0.05_dz0.06}, but consider a significant overall offset redshift error.
    }
    \label{fig:mock_offset}
\end{figure*}
We construct a mock dataset with photo-$z$ errors set to 1\% uniform distribution, 4\% Gaussian distribution with $\sigma_z=0.1(1+z_s)$, and 95\% Gaussian distribution with mean equal to $z_s-0.01$ and $\sigma_z=0.05(1+z_s)$.
Then we consider a scenario that employs the identical binning scheme and noise level as presented in the lower panels of figure~\ref{fig:mock_p_error0.05_dz0.06}.
It should be noted that an overall offset of $\delta_z=0.01$ is typically significantly larger than the current error of photo-$z$ sample, particularly for galaxies with $z_p<1$.
Although this introduces an impact on the reconstruction accuracy, the results presented in figure~\ref{fig:mock_offset} demonstrate that the algorithm is still capable of calibrating redshift distributions.
The degeneracy observed in this scenario is expected to be effectively broken by incorporating spectroscopic data into the calibration process, such as within the $\chi^2$ calculation.

\section{DESI LRG sample with another binning scheme}\label{sec:appendix_LRG_results}
\begin{figure*}
    \centering
	\includegraphics[width=13cm]{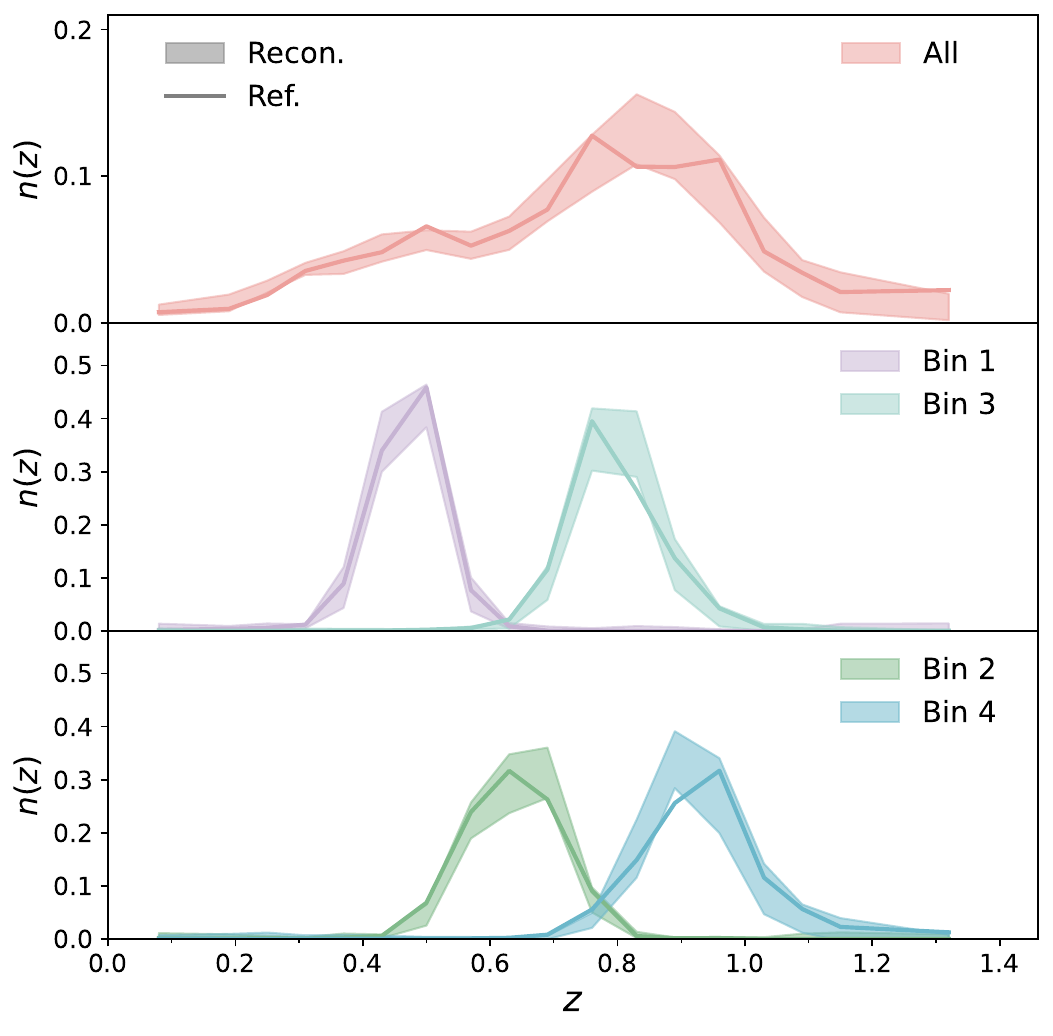}
    \caption{
    Same as figure~\ref{fig:LRG_smaples_nz_dz0.04}, but for another redshift binning scheme with the majority of bin widths set to 0.06.
    }
    \label{fig:LRG_smaples_nz_dz0.06}
\end{figure*}
Figure~\ref{fig:LRG_smaples_nz_dz0.06} shows the reconstructed redshift distributions of DESI LRG samples, with the redshift range divided into 18 bins, the majority of which are set at a width of 0.06.
The bin edges are specified at the following redshift values: 0, 0.16, 0.22, 0.28, 0.34, 0.4, 0.46, 0.54, 0.60, 0.66, 0.72, 0.80, 0.86, 0.92, 1.0, 1.06, 1.12, 1.18, and 1.46.
The reconstructions are consistent with the redshift distributions in figure~\ref{fig:LRG_smaples_nz_dz0.04}, indicating the broad applicability and robustness of the self-calibration algorithm.

\acknowledgments
This work was supported by the National Key R\&D Program of China  (No. 2023YFA1607800, 2023YFA1607802), the National Science Foundation of China (Grant Nos. 12273020, 11621303, 11890691), the China Manned Space Project with Nos. CMS-CSST-2021-A02 and CMS-CSST-2021-A03, the “111” Project of the Ministry of Education under grant No. B20019, and the sponsorship from Yangyang Development Fund.
This work made use of the Gravity Supercomputer at the Department of Astronomy, Shanghai Jiao Tong University.



\bibliographystyle{JHEP}
\bibliography{biblio.bib}

\providecommand{\href}[2]{#2}\begingroup\raggedright\begin{thebibliography}{10}

\bibitem{Peng:2022aa}
H.~{Peng}, H.~{Xu}, L.~{Zhang}, Z.~{Chen} and Y.~{Yu}, \emph{{Self-calibration of photometric redshift scatter from DECaLS DR8 power spectrum and validation with simulated catalogues}}, \href{https://doi.org/10.1093/mnras/stac2713}{\emph{Mon. Not. Roy. Astron. Soc.} {\bfseries 516} (2022) 6210} [\href{https://arxiv.org/abs/2203.14609}{{\ttfamily 2203.14609}}].

\bibitem{Aihara:2018tn}
H.~{Aihara}, N.~{Arimoto}, R.~{Armstrong}, S.~{Arnouts}, N.A.~{Bahcall}, S.~{Bickerton} et~al., \emph{{The Hyper Suprime-Cam SSP Survey: Overview and survey design}}, \href{https://doi.org/10.1093/pasj/psx066}{\emph{Publ. Astron. Soc. Jap.} {\bfseries 70} (2018) S4} [\href{https://arxiv.org/abs/1704.05858}{{\ttfamily 1704.05858}}].

\bibitem{Kuijken:2019vi}
K.~{Kuijken}, C.~{Heymans}, A.~{Dvornik}, H.~{Hildebrandt}, J.T.A.~{de Jong}, A.H.~{Wright} et~al., \emph{{The fourth data release of the Kilo-Degree Survey: ugri imaging and nine-band optical-IR photometry over 1000 square degrees}}, \href{https://doi.org/10.1051/0004-6361/201834918}{\emph{Astron. and Astrophys.} {\bfseries 625} (2019) A2} [\href{https://arxiv.org/abs/1902.11265}{{\ttfamily 1902.11265}}].

\bibitem{Abbott:2018aa}
T.M.C.~{Abbott}, F.B.~{Abdalla}, S.~{Allam}, A.~{Amara}, J.~{Annis}, J.~{Asorey} et~al., \emph{{The Dark Energy Survey: Data Release 1}}, \href{https://doi.org/10.3847/1538-4365/aae9f0}{\emph{Astrophys. J. Suppl.} {\bfseries 239} (2018) 18} [\href{https://arxiv.org/abs/1801.03181}{{\ttfamily 1801.03181}}].

\bibitem{DESI-Collaboration:2016vs}
{DESI Collaboration}, A.~{Aghamousa}, J.~{Aguilar}, S.~{Ahlen}, S.~{Alam}, L.E.~{Allen} et~al., \emph{{The DESI Experiment Part II: Instrument Design}}, {\emph{arXiv e-prints} (2016) arXiv:1611.00037} [\href{https://arxiv.org/abs/1611.00037}{{\ttfamily 1611.00037}}].

\bibitem{DESI-Collaboration:2016vy}
{DESI Collaboration}, A.~{Aghamousa}, J.~{Aguilar}, S.~{Ahlen}, S.~{Alam}, L.E.~{Allen} et~al., \emph{{The DESI Experiment Part I: Science,Targeting, and Survey Design}}, {\emph{arXiv e-prints} (2016) arXiv:1611.00036} [\href{https://arxiv.org/abs/1611.00036}{{\ttfamily 1611.00036}}].

\bibitem{Euclid-Collaboration:2024aa}
{Euclid Collaboration}, Y.~{Mellier}, {Abdurro'uf}, J.A.~{Acevedo Barroso}, A.~{Ach{\'u}carro}, J.~{Adamek} et~al., \emph{{Euclid. I. Overview of the Euclid mission}}, \href{https://doi.org/10.48550/arXiv.2405.13491}{\emph{arXiv e-prints} (2024) arXiv:2405.13491} [\href{https://arxiv.org/abs/2405.13491}{{\ttfamily 2405.13491}}].

\bibitem{Gong:2019tb}
Y.~{Gong}, X.~{Liu}, Y.~{Cao}, X.~{Chen}, Z.~{Fan}, R.~{Li} et~al., \emph{{Cosmology from the Chinese Space Station Optical Survey (CSS-OS)}}, \href{https://doi.org/10.3847/1538-4357/ab391e}{\emph{Astrophys. J.} {\bfseries 883} (2019) 203} [\href{https://arxiv.org/abs/1901.04634}{{\ttfamily 1901.04634}}].

\bibitem{The-LSST-Dark-Energy-Science-Collaboration:2018vo}
{The LSST Dark Energy Science Collaboration}, R.~{Mandelbaum}, T.~{Eifler}, R.~{Hlo{\v{z}}ek}, T.~{Collett}, E.~{Gawiser} et~al., \emph{{The LSST Dark Energy Science Collaboration (DESC) Science Requirements Document}}, {\emph{arXiv e-prints} (2018) arXiv:1809.01669} [\href{https://arxiv.org/abs/1809.01669}{{\ttfamily 1809.01669}}].

\bibitem{Becker:2016aa}
M.R.~{Becker}, M.A.~{Troxel}, N.~{MacCrann}, E.~{Krause}, T.F.~{Eifler}, O.~{Friedrich} et~al., \emph{{Cosmic shear measurements with Dark Energy Survey Science Verification data}}, \href{https://doi.org/10.1103/PhysRevD.94.022002}{\emph{Phys. Rev. D.} {\bfseries 94} (2016) 022002} [\href{https://arxiv.org/abs/1507.05598}{{\ttfamily 1507.05598}}].

\bibitem{Hildebrandt:2021vt}
H.~{Hildebrandt}, J.L.~{van den Busch}, A.H.~{Wright}, C.~{Blake}, B.~{Joachimi}, K.~{Kuijken} et~al., \emph{{KiDS-1000 catalogue: Redshift distributions and their calibration}}, \href{https://doi.org/10.1051/0004-6361/202039018}{\emph{Astron. and Astrophys.} {\bfseries 647} (2021) A124} [\href{https://arxiv.org/abs/2007.15635}{{\ttfamily 2007.15635}}].

\bibitem{Myles:2021aa}
J.~{Myles}, A.~{Alarcon}, A.~{Amon}, C.~{S{\'a}nchez}, S.~{Everett}, J.~{DeRose} et~al., \emph{{Dark Energy Survey Year 3 results: redshift calibration of the weak lensing source galaxies}}, \href{https://doi.org/10.1093/mnras/stab1515}{\emph{Mon. Not. Roy. Astron. Soc.} {\bfseries 505} (2021) 4249} [\href{https://arxiv.org/abs/2012.08566}{{\ttfamily 2012.08566}}].

\bibitem{Mandelbaum:2018aa}
R.~{Mandelbaum}, \emph{{Weak Lensing for Precision Cosmology}}, \href{https://doi.org/10.1146/annurev-astro-081817-051928}{\emph{Annual Review of Astronomy and Astrophysics.} {\bfseries 56} (2018) 393} [\href{https://arxiv.org/abs/1710.03235}{{\ttfamily 1710.03235}}].

\bibitem{Yao:2024aa}
J.~{Yao}, H.~{Shan}, R.~{Li}, Y.~{Xu}, D.~{Fan}, D.~{Liu} et~al., \emph{{CSST WL preparation I: forecast the impact from non-Gaussian covariances and requirements on systematics control}}, \href{https://doi.org/10.1093/mnras/stad3563}{\emph{Mon. Not. Roy. Astron. Soc.} {\bfseries 527} (2024) 5206} [\href{https://arxiv.org/abs/2304.04489}{{\ttfamily 2304.04489}}].

\bibitem{Bernstein:2010wc}
G.~{Bernstein} and D.~{Huterer}, \emph{{Catastrophic photometric redshift errors: weak-lensing survey requirements}}, \href{https://doi.org/10.1111/j.1365-2966.2009.15748.x}{\emph{Mon. Not. Roy. Astron. Soc.} {\bfseries 401} (2010) 1399} [\href{https://arxiv.org/abs/0902.2782}{{\ttfamily 0902.2782}}].

\bibitem{Salvato:2019aa}
M.~{Salvato}, O.~{Ilbert} and B.~{Hoyle}, \emph{{The many flavours of photometric redshifts}}, \href{https://doi.org/10.1038/s41550-018-0478-0}{\emph{Nature Astronomy} {\bfseries 3} (2019) 212} [\href{https://arxiv.org/abs/1805.12574}{{\ttfamily 1805.12574}}].

\bibitem{Newman:2022vy}
J.A.~{Newman} and D.~{Gruen}, \emph{{Photometric Redshifts for Next-Generation Surveys}}, \href{https://doi.org/10.1146/annurev-astro-032122-014611}{\emph{Annual Review of Astronomy and Astrophysics.} {\bfseries 60} (2022) 363} [\href{https://arxiv.org/abs/2206.13633}{{\ttfamily 2206.13633}}].

\bibitem{Lima:2008vl}
M.~{Lima}, C.E.~{Cunha}, H.~{Oyaizu}, J.~{Frieman}, H.~{Lin} and E.S.~{Sheldon}, \emph{{Estimating the redshift distribution of photometric galaxy samples}}, \href{https://doi.org/10.1111/j.1365-2966.2008.13510.x}{\emph{Mon. Not. Roy. Astron. Soc.} {\bfseries 390} (2008) 118} [\href{https://arxiv.org/abs/0801.3822}{{\ttfamily 0801.3822}}].

\bibitem{Bonnett:2016wb}
C.~{Bonnett}, M.A.~{Troxel}, W.~{Hartley}, A.~{Amara}, B.~{Leistedt}, M.R.~{Becker} et~al., \emph{{Redshift distributions of galaxies in the Dark Energy Survey Science Verification shear catalogue and implications for weak lensing}}, \href{https://doi.org/10.1103/PhysRevD.94.042005}{\emph{Phys. Rev. D.} {\bfseries 94} (2016) 042005} [\href{https://arxiv.org/abs/1507.05909}{{\ttfamily 1507.05909}}].

\bibitem{Hildebrandt:2017ua}
H.~{Hildebrandt}, M.~{Viola}, C.~{Heymans}, S.~{Joudaki}, K.~{Kuijken}, C.~{Blake} et~al., \emph{{KiDS-450: cosmological parameter constraints from tomographic weak gravitational lensing}}, \href{https://doi.org/10.1093/mnras/stw2805}{\emph{Mon. Not. Roy. Astron. Soc.} {\bfseries 465} (2017) 1454} [\href{https://arxiv.org/abs/1606.05338}{{\ttfamily 1606.05338}}].

\bibitem{Wright:2019aa}
A.H.~{Wright}, H.~{Hildebrandt}, K.~{Kuijken}, T.~{Erben}, R.~{Blake}, H.~{Buddelmeijer} et~al., \emph{{KiDS+VIKING-450: A new combined optical and near-infrared dataset for cosmology and astrophysics}}, \href{https://doi.org/10.1051/0004-6361/201834879}{\emph{Astron. and Astrophys.} {\bfseries 632} (2019) A34} [\href{https://arxiv.org/abs/1812.06077}{{\ttfamily 1812.06077}}].

\bibitem{Hildebrandt:2020aa}
H.~{Hildebrandt}, F.~{K{\"o}hlinger}, J.L.~{van den Busch}, B.~{Joachimi}, C.~{Heymans}, A.~{Kannawadi} et~al., \emph{{KiDS+VIKING-450: Cosmic shear tomography with optical and infrared data}}, \href{https://doi.org/10.1051/0004-6361/201834878}{\emph{Astron. and Astrophys.} {\bfseries 633} (2020) A69} [\href{https://arxiv.org/abs/1812.06076}{{\ttfamily 1812.06076}}].

\bibitem{Kohonen1982}
T.~Kohonen, \emph{Self-organized formation of topologically correct feature maps}, {\emph{Biol. Cybern.} {\bfseries 43} (1982) 59}.

\bibitem{Masters:2015wp}
D.~{Masters}, P.~{Capak}, D.~{Stern}, O.~{Ilbert}, M.~{Salvato}, S.~{Schmidt} et~al., \emph{{Mapping the Galaxy Color-Redshift Relation: Optimal Photometric Redshift Calibration Strategies for Cosmology Surveys}}, \href{https://doi.org/10.1088/0004-637X/813/1/53}{\emph{Astrophys. J.} {\bfseries 813} (2015) 53} [\href{https://arxiv.org/abs/1509.03318}{{\ttfamily 1509.03318}}].

\bibitem{Herbel:2017aa}
J.~{Herbel}, T.~{Kacprzak}, A.~{Amara}, A.~{Refregier}, C.~{Bruderer} and A.~{Nicola}, \emph{{The redshift distribution of cosmological samples: a forward modeling approach}}, \href{https://doi.org/10.1088/1475-7516/2017/08/035}{\emph{JCAP.} {\bfseries 2017} (2017) 035} [\href{https://arxiv.org/abs/1705.05386}{{\ttfamily 1705.05386}}].

\bibitem{Buchs:2019tx}
R.~{Buchs}, C.~{Davis}, D.~{Gruen}, J.~{DeRose}, A.~{Alarcon}, G.M.~{Bernstein} et~al., \emph{{Phenotypic redshifts with self-organizing maps: A novel method to characterize redshift distributions of source galaxies for weak lensing}}, \href{https://doi.org/10.1093/mnras/stz2162}{\emph{Mon. Not. Roy. Astron. Soc.} {\bfseries 489} (2019) 820} [\href{https://arxiv.org/abs/1901.05005}{{\ttfamily 1901.05005}}].

\bibitem{Wright:2020vs}
A.H.~{Wright}, H.~{Hildebrandt}, J.L.~{van den Busch} and C.~{Heymans}, \emph{{Photometric redshift calibration with self-organising maps}}, \href{https://doi.org/10.1051/0004-6361/201936782}{\emph{Astron. and Astrophys.} {\bfseries 637} (2020) A100} [\href{https://arxiv.org/abs/1909.09632}{{\ttfamily 1909.09632}}].

\bibitem{Giannini:2024aa}
G.~{Giannini}, A.~{Alarcon}, M.~{Gatti}, A.~{Porredon}, M.~{Crocce}, G.M.~{Bernstein} et~al., \emph{{Dark Energy Survey Year 3 results: redshift calibration of the MAGLIM lens sample from the combination of SOMPZ and clustering and its impact on cosmology}}, \href{https://doi.org/10.1093/mnras/stad2945}{\emph{Mon. Not. Roy. Astron. Soc.} {\bfseries 527} (2024) 2010} [\href{https://arxiv.org/abs/2209.05853}{{\ttfamily 2209.05853}}].

\bibitem{Newman:2008vv}
J.A.~{Newman}, \emph{{Calibrating Redshift Distributions beyond Spectroscopic Limits with Cross-Correlations}}, \href{https://doi.org/10.1086/589982}{\emph{Astrophys. J.} {\bfseries 684} (2008) 88} [\href{https://arxiv.org/abs/0805.1409}{{\ttfamily 0805.1409}}].

\bibitem{Matthews:2010un}
D.J.~{Matthews} and J.A.~{Newman}, \emph{{Reconstructing Redshift Distributions with Cross-correlations: Tests and an Optimized Recipe}}, \href{https://doi.org/10.1088/0004-637X/721/1/456}{\emph{Astrophys. J.} {\bfseries 721} (2010) 456} [\href{https://arxiv.org/abs/1003.0687}{{\ttfamily 1003.0687}}].

\bibitem{McQuinn:2013uh}
M.~{McQuinn} and M.~{White}, \emph{{On using angular cross-correlations to determine source redshift distributions}}, \href{https://doi.org/10.1093/mnras/stt914}{\emph{Mon. Not. Roy. Astron. Soc.} {\bfseries 433} (2013) 2857} [\href{https://arxiv.org/abs/1302.0857}{{\ttfamily 1302.0857}}].

\bibitem{Schmidt:2013va}
S.J.~{Schmidt}, B.~{M{\'e}nard}, R.~{Scranton}, C.~{Morrison} and C.K.~{McBride}, \emph{{Recovering redshift distributions with cross-correlations: pushing the boundaries}}, \href{https://doi.org/10.1093/mnras/stt410}{\emph{Mon. Not. Roy. Astron. Soc.} {\bfseries 431} (2013) 3307} [\href{https://arxiv.org/abs/1303.0292}{{\ttfamily 1303.0292}}].

\bibitem{Choi:2016aa}
A.~{Choi}, C.~{Heymans}, C.~{Blake}, H.~{Hildebrandt}, C.A.J.~{Duncan}, T.~{Erben} et~al., \emph{{CFHTLenS and RCSLenS: testing photometric redshift distributions using angular cross-correlations with spectroscopic galaxy surveys}}, \href{https://doi.org/10.1093/mnras/stw2241}{\emph{Mon. Not. Roy. Astron. Soc.} {\bfseries 463} (2016) 3737} [\href{https://arxiv.org/abs/1512.03626}{{\ttfamily 1512.03626}}].

\bibitem{McLeod:2017vz}
M.~{McLeod}, S.T.~{Balan} and F.B.~{Abdalla}, \emph{{A joint analysis for cosmology and photometric redshift calibration using cross-correlations}}, \href{https://doi.org/10.1093/mnras/stw2989}{\emph{Mon. Not. Roy. Astron. Soc.} {\bfseries 466} (2017) 3558} [\href{https://arxiv.org/abs/1612.00307}{{\ttfamily 1612.00307}}].

\bibitem{Morrison:2017aa}
C.B.~{Morrison}, H.~{Hildebrandt}, S.J.~{Schmidt}, I.K.~{Baldry}, M.~{Bilicki}, A.~{Choi} et~al., \emph{{the-wizz: clustering redshift estimation for everyone}}, \href{https://doi.org/10.1093/mnras/stx342}{\emph{Mon. Not. Roy. Astron. Soc.} {\bfseries 467} (2017) 3576} [\href{https://arxiv.org/abs/1609.09085}{{\ttfamily 1609.09085}}].

\bibitem{Gatti:2018tg}
M.~{Gatti}, P.~{Vielzeuf}, C.~{Davis}, R.~{Cawthon}, M.M.~{Rau}, J.~{DeRose} et~al., \emph{{Dark Energy Survey Year 1 results: cross-correlation redshifts - methods and systematics characterization}}, \href{https://doi.org/10.1093/mnras/sty466}{\emph{Mon. Not. Roy. Astron. Soc.} {\bfseries 477} (2018) 1664} [\href{https://arxiv.org/abs/1709.00992}{{\ttfamily 1709.00992}}].

\bibitem{Cawthon:2022aa}
R.~{Cawthon}, J.~{Elvin-Poole}, A.~{Porredon}, M.~{Crocce}, G.~{Giannini}, M.~{Gatti} et~al., \emph{{Dark Energy Survey Year 3 results: calibration of lens sample redshift distributions using clustering redshifts with BOSS/eBOSS}}, \href{https://doi.org/10.1093/mnras/stac1160}{\emph{Mon. Not. Roy. Astron. Soc.} {\bfseries 513} (2022) 5517} [\href{https://arxiv.org/abs/2012.12826}{{\ttfamily 2012.12826}}].

\bibitem{Gatti:2022va}
M.~{Gatti}, G.~{Giannini}, G.M.~{Bernstein}, A.~{Alarcon}, J.~{Myles}, A.~{Amon} et~al., \emph{{Dark Energy Survey Year 3 Results: clustering redshifts - calibration of the weak lensing source redshift distributions with redMaGiC and BOSS/eBOSS}}, \href{https://doi.org/10.1093/mnras/stab3311}{\emph{Mon. Not. Roy. Astron. Soc.} {\bfseries 510} (2022) 1223} [\href{https://arxiv.org/abs/2012.08569}{{\ttfamily 2012.08569}}].

\bibitem{Rau:2022ud}
M.M.~{Rau}, C.B.~{Morrison}, S.J.~{Schmidt}, S.~{Wilson}, R.~{Mandelbaum}, Y.Y.~{Mao} et~al., \emph{{A composite likelihood approach for inference under photometric redshift uncertainty}}, \href{https://doi.org/10.1093/mnras/stab3290}{\emph{Mon. Not. Roy. Astron. Soc.} {\bfseries 509} (2022) 4886} [\href{https://arxiv.org/abs/2101.01184}{{\ttfamily 2101.01184}}].

\bibitem{Stolzner:2023aa}
B.~{St{\"o}lzner}, B.~{Joachimi}, A.~{Korn} and {LSST Dark Energy Science Collaboration}, \emph{{Optimizing the shape of photometric redshift distributions with clustering cross-correlations}}, \href{https://doi.org/10.1093/mnras/stac3630}{\emph{Mon. Not. Roy. Astron. Soc.} {\bfseries 519} (2023) 2438} [\href{https://arxiv.org/abs/2205.13622}{{\ttfamily 2205.13622}}].

\bibitem{Leistedt:2016aa}
B.~{Leistedt}, D.J.~{Mortlock} and H.V.~{Peiris}, \emph{{Hierarchical Bayesian inference of galaxy redshift distributions from photometric surveys}}, \href{https://doi.org/10.1093/mnras/stw1304}{\emph{Mon. Not. Roy. Astron. Soc.} {\bfseries 460} (2016) 4258} [\href{https://arxiv.org/abs/1602.05960}{{\ttfamily 1602.05960}}].

\bibitem{Leistedt:2019aa}
B.~{Leistedt}, D.W.~{Hogg}, R.H.~{Wechsler} and J.~{DeRose}, \emph{{Hierarchical Modeling and Statistical Calibration for Photometric Redshifts}}, \href{https://doi.org/10.3847/1538-4357/ab2d29}{\emph{Astrophys. J.} {\bfseries 881} (2019) 80} [\href{https://arxiv.org/abs/1807.01391}{{\ttfamily 1807.01391}}].

\bibitem{Sanchez:2019aa}
C.~{S{\'a}nchez} and G.M.~{Bernstein}, \emph{{Redshift inference from the combination of galaxy colours and clustering in a hierarchical Bayesian model}}, \href{https://doi.org/10.1093/mnras/sty3222}{\emph{Mon. Not. Roy. Astron. Soc.} {\bfseries 483} (2019) 2801} [\href{https://arxiv.org/abs/1807.11873}{{\ttfamily 1807.11873}}].

\bibitem{Alarcon:2020aa}
A.~{Alarcon}, C.~{S{\'a}nchez}, G.M.~{Bernstein} and E.~{Gazta{\~n}aga}, \emph{{Redshift inference from the combination of galaxy colours and clustering in a hierarchical Bayesian model - Application to realistic N-body simulations}}, \href{https://doi.org/10.1093/mnras/staa2478}{\emph{Mon. Not. Roy. Astron. Soc.} {\bfseries 498} (2020) 2614} [\href{https://arxiv.org/abs/1910.07127}{{\ttfamily 1910.07127}}].

\bibitem{Malz:2022aa}
A.I.~{Malz} and D.W.~{Hogg}, \emph{{How to Obtain the Redshift Distribution from Probabilistic Redshift Estimates}}, \href{https://doi.org/10.3847/1538-4357/ac062f}{\emph{Astrophys. J.} {\bfseries 928} (2022) 127}.

\bibitem{Autenrieth:2024aa}
M.~{Autenrieth}, A.H.~{Wright}, R.~{Trotta}, D.A.~{van Dyk}, D.C.~{Stenning} and B.~{Joachimi}, \emph{{Improved Weak Lensing Photometric Redshift Calibration via StratLearn and Hierarchical Modeling}}, \href{https://doi.org/10.48550/arXiv.2401.04687}{\emph{arXiv e-prints} (2024) arXiv:2401.04687} [\href{https://arxiv.org/abs/2401.04687}{{\ttfamily 2401.04687}}].

\bibitem{Gruen:2017aa}
D.~{Gruen} and F.~{Brimioulle}, \emph{{Selection biases in empirical p(z) methods for weak lensing}}, \href{https://doi.org/10.1093/mnras/stx471}{\emph{Mon. Not. Roy. Astron. Soc.} {\bfseries 468} (2017) 769} [\href{https://arxiv.org/abs/1610.01160}{{\ttfamily 1610.01160}}].

\bibitem{Hemmati:2019aa}
S.~{Hemmati}, P.~{Capak}, D.~{Masters}, I.~{Davidzon}, O.~{Dor{\`e}}, J.~{Kruk} et~al., \emph{{Photometric Redshift Calibration Requirements for WFIRST Weak-lensing Cosmology: Predictions from CANDELS}}, \href{https://doi.org/10.3847/1538-4357/ab1be5}{\emph{Astrophys. J.} {\bfseries 877} (2019) 117} [\href{https://arxiv.org/abs/1808.10458}{{\ttfamily 1808.10458}}].

\bibitem{Hartley:2020aa}
W.G.~{Hartley}, C.~{Chang}, S.~{Samani}, A.~{Carnero Rosell}, T.M.~{Davis}, B.~{Hoyle} et~al., \emph{{The impact of spectroscopic incompleteness in direct calibration of redshift distributions for weak lensing surveys}}, \href{https://doi.org/10.1093/mnras/staa1812}{\emph{Mon. Not. Roy. Astron. Soc.} {\bfseries 496} (2020) 4769} [\href{https://arxiv.org/abs/2003.10454}{{\ttfamily 2003.10454}}].

\bibitem{Sanchez:2020aa}
C.~{S{\'a}nchez}, M.~{Raveri}, A.~{Alarcon} and G.M.~{Bernstein}, \emph{{Propagating sample variance uncertainties in redshift calibration: simulations, theory, and application to the COSMOS2015 data}}, \href{https://doi.org/10.1093/mnras/staa2542}{\emph{Mon. Not. Roy. Astron. Soc.} {\bfseries 498} (2020) 2984} [\href{https://arxiv.org/abs/2004.09542}{{\ttfamily 2004.09542}}].

\bibitem{Schneider:2006ta}
M.~{Schneider}, L.~{Knox}, H.~{Zhan} and A.~{Connolly}, \emph{{Using Galaxy Two-Point Correlation Functions to Determine the Redshift Distributions of Galaxies Binned by Photometric Redshift}}, \href{https://doi.org/10.1086/507675}{\emph{Astrophys. J.} {\bfseries 651} (2006) 14} [\href{https://arxiv.org/abs/astro-ph/0606098}{{\ttfamily astro-ph/0606098}}].

\bibitem{Benjamin:2010aa}
J.~{Benjamin}, L.~{van Waerbeke}, B.~{M{\'e}nard} and M.~{Kilbinger}, \emph{{Photometric redshifts: estimating their contamination and distribution using clustering information}}, \href{https://doi.org/10.1111/j.1365-2966.2010.17191.x}{\emph{Mon. Not. Roy. Astron. Soc.} {\bfseries 408} (2010) 1168} [\href{https://arxiv.org/abs/1002.2266}{{\ttfamily 1002.2266}}].

\bibitem{Zhang:2010wr}
P.~{Zhang}, U.-L.~{Pen} and G.~{Bernstein}, \emph{{Self-calibration of photometric redshift scatter in weak-lensing surveys}}, \href{https://doi.org/10.1111/j.1365-2966.2010.16445.x}{\emph{Mon. Not. Roy. Astron. Soc.} {\bfseries 405} (2010) 359} [\href{https://arxiv.org/abs/0910.4181}{{\ttfamily 0910.4181}}].

\bibitem{Zhang:2017um}
L.~{Zhang}, Y.~{Yu} and P.~{Zhang}, \emph{{Non-negative Matrix Factorization for Self-calibration of Photometric Redshift Scatter in Weak-lensing Surveys}}, \href{https://doi.org/10.3847/1538-4357/aa8c72}{\emph{Astrophys. J.} {\bfseries 848} (2017) 44} [\href{https://arxiv.org/abs/1612.04042}{{\ttfamily 1612.04042}}].

\bibitem{Schaan:2020up}
E.~{Schaan}, S.~{Ferraro} and U.~{Seljak}, \emph{{Photo-z outlier self-calibration in weak lensing surveys}}, \href{https://doi.org/10.1088/1475-7516/2020/12/001}{\emph{JCAP.} {\bfseries 2020} (2020) 001} [\href{https://arxiv.org/abs/2007.12795}{{\ttfamily 2007.12795}}].

\bibitem{Pyne:2021aa}
S.~{Pyne} and B.~{Joachimi}, \emph{{Self-calibration of weak lensing systematic effects using combined two- and three-point statistics}}, \href{https://doi.org/10.1093/mnras/stab413}{\emph{Mon. Not. Roy. Astron. Soc.} {\bfseries 503} (2021) 2300} [\href{https://arxiv.org/abs/2010.00614}{{\ttfamily 2010.00614}}].

\bibitem{Stolzner:2021va}
B.~{St{\"o}lzner}, B.~{Joachimi}, A.~{Korn}, H.~{Hildebrandt} and A.H.~{Wright}, \emph{{Self-calibration and robust propagation of photometric redshift distribution uncertainties in weak gravitational lensing}}, \href{https://doi.org/10.1051/0004-6361/202040130}{\emph{Astron. and Astrophys.} {\bfseries 650} (2021) A148} [\href{https://arxiv.org/abs/2012.07707}{{\ttfamily 2012.07707}}].

\bibitem{Xu:2023aa}
H.~{Xu}, P.~{Zhang}, H.~{Peng}, Y.~{Yu}, L.~{Zhang}, J.~{Yao} et~al., \emph{{Using angular two-point correlations to self-calibrate the photometric redshift distributions of DECaLS DR9}}, \href{https://doi.org/10.1093/mnras/stad136}{\emph{Mon. Not. Roy. Astron. Soc.} {\bfseries 520} (2023) 161} [\href{https://arxiv.org/abs/2209.03967}{{\ttfamily 2209.03967}}].

\bibitem{Song:2024aa}
R.~{Song}, K.C.~{Chan}, H.~{Xu} and W.~{Zheng}, \emph{{Measurement of the photometric baryon acoustic oscillations with self-calibrated redshift distribution}}, \href{https://doi.org/10.1093/mnras/stae910}{\emph{Mon. Not. Roy. Astron. Soc.} {\bfseries 530} (2024) 881} [\href{https://arxiv.org/abs/2402.18827}{{\ttfamily 2402.18827}}].

\bibitem{Alsing:2023aa}
J.~{Alsing}, H.~{Peiris}, D.~{Mortlock}, J.~{Leja} and B.~{Leistedt}, \emph{{Forward Modeling of Galaxy Populations for Cosmological Redshift Distribution Inference}}, \href{https://doi.org/10.3847/1538-4365/ac9583}{\emph{Astrophys. J. Suppl.} {\bfseries 264} (2023) 29} [\href{https://arxiv.org/abs/2207.05819}{{\ttfamily 2207.05819}}].

\bibitem{Leistedt:2023aa}
B.~{Leistedt}, J.~{Alsing}, H.~{Peiris}, D.~{Mortlock} and J.~{Leja}, \emph{{Hierarchical Bayesian Inference of Photometric Redshifts with Stellar Population Synthesis Models}}, \href{https://doi.org/10.3847/1538-4365/ac9d99}{\emph{Astrophys. J. Suppl.} {\bfseries 264} (2023) 23} [\href{https://arxiv.org/abs/2207.07673}{{\ttfamily 2207.07673}}].

\bibitem{Paatero:1994aa}
P.~{Paatero} and U.~{Tapper}, \emph{{Positive matrix factorization: A non-negative factor model with optimal utilization of error estimates of data values}}, \href{https://doi.org/10.1002/env.3170050203}{\emph{Environmetrics} {\bfseries 5} (1994) 111}.

\bibitem{Lee:2001aa}
D.~Lee and H.S.~Seung, \emph{Algorithms for non-negative matrix factorization}, {\emph{Advances in Neural Information Processing Systems} {\bfseries 13} (2000) }.

\bibitem{Tsalmantza:2012aa}
P.~{Tsalmantza} and D.W.~{Hogg}, \emph{{A Data-driven Model for Spectra: Finding Double Redshifts in the Sloan Digital Sky Survey}}, \href{https://doi.org/10.1088/0004-637X/753/2/122}{\emph{Astrophys. J.} {\bfseries 753} (2012) 122} [\href{https://arxiv.org/abs/1201.3370}{{\ttfamily 1201.3370}}].

\bibitem{Zhu:2016aa}
G.~{Zhu}, \emph{{Nonnegative Matrix Factorization (NMF) with Heteroscedastic Uncertainties and Missing data}}, \href{https://doi.org/10.48550/arXiv.1612.06037}{\emph{arXiv e-prints} (2016) arXiv:1612.06037} [\href{https://arxiv.org/abs/1612.06037}{{\ttfamily 1612.06037}}].

\bibitem{Ren:2018aa}
B.~{Ren}, L.~{Pueyo}, G.B.~{Zhu}, J.~{Debes} and G.~{Duch{\^e}ne}, \emph{{Non-negative Matrix Factorization: Robust Extraction of Extended Structures}}, \href{https://doi.org/10.3847/1538-4357/aaa1f2}{\emph{Astrophys. J.} {\bfseries 852} (2018) 104} [\href{https://arxiv.org/abs/1712.10317}{{\ttfamily 1712.10317}}].

\bibitem{Green:2023aa}
D.~{Green} and S.~{Bailey}, \emph{{Algorithms for Non-Negative Matrix Factorization on Noisy Data With Negative Values}}, \href{https://doi.org/10.48550/arXiv.2311.04855}{\emph{arXiv e-prints} (2023) arXiv:2311.04855} [\href{https://arxiv.org/abs/2311.04855}{{\ttfamily 2311.04855}}].

\bibitem{Zhu:2013aa}
Z.~Zhu, Z.~Yang and E.~Oja, \emph{Multiplicative updates for learning with stochastic matrices}, {\emph{Scandinavian Conference on Image Analysis} (2013) 143}.

\bibitem{Zhu:2010aa}
Z.~Yang and E.~Oja, \emph{Linear and nonlinear projective nonnegative matrix factorization}, \href{https://doi.org/10.1109/TNN.2010.2041361}{\emph{IEEE Transactions on Neural Networks} {\bfseries 21} (2010) 734}.

\bibitem{Zhou:2023aa}
R.~{Zhou}, S.~{Ferraro}, M.~{White}, J.~{DeRose}, N.~{Sailer}, J.~{Aguilar} et~al., \emph{{DESI luminous red galaxy samples for cross-correlations}}, \href{https://doi.org/10.1088/1475-7516/2023/11/097}{\emph{JCAP.} {\bfseries 2023} (2023) 097} [\href{https://arxiv.org/abs/2309.06443}{{\ttfamily 2309.06443}}].

\bibitem{DESI-Collaboration:2022aa}
{DESI Collaboration}, B.~{Abareshi}, J.~{Aguilar}, S.~{Ahlen}, S.~{Alam}, D.M.~{Alexander} et~al., \emph{{Overview of the Instrumentation for the Dark Energy Spectroscopic Instrument}}, \href{https://doi.org/10.3847/1538-3881/ac882b}{\emph{Astron. J.} {\bfseries 164} (2022) 207} [\href{https://arxiv.org/abs/2205.10939}{{\ttfamily 2205.10939}}].

\bibitem{DESI-Collaboration:2024ab}
{DESI Collaboration}, A.G.~{Adame}, J.~{Aguilar}, S.~{Ahlen}, S.~{Alam}, D.M.~{Alexander} et~al., \emph{{DESI 2024 VI: Cosmological Constraints from the Measurements of Baryon Acoustic Oscillations}}, \href{https://doi.org/10.48550/arXiv.2404.03002}{\emph{arXiv e-prints} (2024) arXiv:2404.03002} [\href{https://arxiv.org/abs/2404.03002}{{\ttfamily 2404.03002}}].

\bibitem{Myers:2023aa}
A.D.~{Myers}, J.~{Moustakas}, S.~{Bailey}, B.A.~{Weaver}, A.P.~{Cooper}, J.E.~{Forero-Romero} et~al., \emph{{The Target-selection Pipeline for the Dark Energy Spectroscopic Instrument}}, \href{https://doi.org/10.3847/1538-3881/aca5f9}{\emph{Astron. J.} {\bfseries 165} (2023) 50} [\href{https://arxiv.org/abs/2208.08518}{{\ttfamily 2208.08518}}].

\bibitem{Hahn:2023ab}
C.~{Hahn}, M.J.~{Wilson}, O.~{Ruiz-Macias}, S.~{Cole}, D.H.~{Weinberg}, J.~{Moustakas} et~al., \emph{{The DESI Bright Galaxy Survey: Final Target Selection, Design, and Validation}}, \href{https://doi.org/10.3847/1538-3881/accff8}{\emph{Astron. J.} {\bfseries 165} (2023) 253} [\href{https://arxiv.org/abs/2208.08512}{{\ttfamily 2208.08512}}].

\bibitem{Zhou:2023ab}
R.~{Zhou}, B.~{Dey}, J.A.~{Newman}, D.J.~{Eisenstein}, K.~{Dawson}, S.~{Bailey} et~al., \emph{{Target Selection and Validation of DESI Luminous Red Galaxies}}, \href{https://doi.org/10.3847/1538-3881/aca5fb}{\emph{Astron. J.} {\bfseries 165} (2023) 58} [\href{https://arxiv.org/abs/2208.08515}{{\ttfamily 2208.08515}}].

\bibitem{Raichoor:2023aa}
A.~{Raichoor}, J.~{Moustakas}, J.A.~{Newman}, T.~{Karim}, S.~{Ahlen}, S.~{Alam} et~al., \emph{{Target Selection and Validation of DESI Emission Line Galaxies}}, \href{https://doi.org/10.3847/1538-3881/acb213}{\emph{Astron. J.} {\bfseries 165} (2023) 126} [\href{https://arxiv.org/abs/2208.08513}{{\ttfamily 2208.08513}}].

\bibitem{Chaussidon:2023aa}
E.~{Chaussidon}, C.~{Y{\`e}che}, N.~{Palanque-Delabrouille}, D.M.~{Alexander}, J.~{Yang}, S.~{Ahlen} et~al., \emph{{Target Selection and Validation of DESI Quasars}}, \href{https://doi.org/10.3847/1538-4357/acb3c2}{\emph{Astrophys. J.} {\bfseries 944} (2023) 107} [\href{https://arxiv.org/abs/2208.08511}{{\ttfamily 2208.08511}}].

\bibitem{Dey:2019aa}
A.~{Dey}, D.J.~{Schlegel}, D.~{Lang}, R.~{Blum}, K.~{Burleigh}, X.~{Fan} et~al., \emph{{Overview of the DESI Legacy Imaging Surveys}}, \href{https://doi.org/10.3847/1538-3881/ab089d}{\emph{Astron. J.} {\bfseries 157} (2019) 168} [\href{https://arxiv.org/abs/1804.08657}{{\ttfamily 1804.08657}}].

\bibitem{DESI-Collaboration:2024ad}
{DESI Collaboration}, A.G.~{Adame}, J.~{Aguilar}, S.~{Ahlen}, S.~{Alam}, G.~{Aldering} et~al., \emph{{The Early Data Release of the Dark Energy Spectroscopic Instrument}}, \href{https://doi.org/10.3847/1538-3881/ad3217}{\emph{Astron. J.} {\bfseries 168} (2024) 58} [\href{https://arxiv.org/abs/2306.06308}{{\ttfamily 2306.06308}}].

\bibitem{DESI-Collaboration:2024aa}
{DESI Collaboration}, A.G.~{Adame}, J.~{Aguilar}, S.~{Ahlen}, S.~{Alam}, G.~{Aldering} et~al., \emph{{Validation of the Scientific Program for the Dark Energy Spectroscopic Instrument}}, \href{https://doi.org/10.3847/1538-3881/ad0b08}{\emph{Astron. J.} {\bfseries 167} (2024) 62} [\href{https://arxiv.org/abs/2306.06307}{{\ttfamily 2306.06307}}].

\bibitem{Xu:2023ab}
H.~{Xu}, H.~{Li}, J.~{Zhang}, X.~{Yang}, P.~{Zhang}, M.~{He} et~al., \emph{{DESI Legacy Imaging Surveys Data Release 9: Cosmological constraints from galaxy clustering and weak lensing using the minimal bias model}}, \href{https://doi.org/10.1007/s11433-023-2242-8}{\emph{Science China Physics, Mechanics, and Astronomy} {\bfseries 66} (2023) 129811} [\href{https://arxiv.org/abs/2310.03066}{{\ttfamily 2310.03066}}].

\bibitem{Planck-Collaboration:2020aa}
{Planck Collaboration}, N.~{Aghanim}, Y.~{Akrami}, M.~{Ashdown}, J.~{Aumont}, C.~{Baccigalupi} et~al., \emph{{Planck 2018 results. VI. Cosmological parameters}}, \href{https://doi.org/10.1051/0004-6361/201833910}{\emph{Astron. and Astrophys.} {\bfseries 641} (2020) A6} [\href{https://arxiv.org/abs/1807.06209}{{\ttfamily 1807.06209}}].

\bibitem{Chisari:2019wv}
N.E.~{Chisari}, D.~{Alonso}, E.~{Krause}, C.D.~{Leonard}, P.~{Bull}, J.~{Neveu} et~al., \emph{{Core Cosmology Library: Precision Cosmological Predictions for LSST}}, \href{https://doi.org/10.3847/1538-4365/ab1658}{\emph{Astrophys. J. Suppl.} {\bfseries 242} (2019) 2} [\href{https://arxiv.org/abs/1812.05995}{{\ttfamily 1812.05995}}].

\bibitem{Kilbinger:2017aa}
M.~{Kilbinger}, C.~{Heymans}, M.~{Asgari}, S.~{Joudaki}, P.~{Schneider}, P.~{Simon} et~al., \emph{{Precision calculations of the cosmic shear power spectrum projection}}, \href{https://doi.org/10.1093/mnras/stx2082}{\emph{Mon. Not. Roy. Astron. Soc.} {\bfseries 472} (2017) 2126} [\href{https://arxiv.org/abs/1702.05301}{{\ttfamily 1702.05301}}].

\bibitem{Leonard:2023aa}
C.D.~{Leonard}, T.~{Ferreira}, X.~{Fang}, R.~{Reischke}, N.~{Schoeneberg}, T.~{Tr{\"o}ster} et~al., \emph{{The N5K Challenge: Non-Limber Integration for LSST Cosmology}}, \href{https://doi.org/10.21105/astro.2212.04291}{\emph{The Open Journal of Astrophysics} {\bfseries 6} (2023) 8} [\href{https://arxiv.org/abs/2212.04291}{{\ttfamily 2212.04291}}].

\bibitem{Gorski:2005te}
K.M.~{G{\'o}rski}, E.~{Hivon}, A.J.~{Banday}, B.D.~{Wandelt}, F.K.~{Hansen}, M.~{Reinecke} et~al., \emph{{HEALPix: A Framework for High-Resolution Discretization and Fast Analysis of Data Distributed on the Sphere}}, \href{https://doi.org/10.1086/427976}{\emph{Astrophys. J.} {\bfseries 622} (2005) 759} [\href{https://arxiv.org/abs/astro-ph/0409513}{{\ttfamily astro-ph/0409513}}].

\bibitem{Alonso:2019aa}
D.~{Alonso}, J.~{Sanchez}, A.~{Slosar} and {LSST Dark Energy Science Collaboration}, \emph{{A unified pseudo-C$_{{\ensuremath{\ell}}}$ framework}}, \href{https://doi.org/10.1093/mnras/stz093}{\emph{Mon. Not. Roy. Astron. Soc.} {\bfseries 484} (2019) 4127} [\href{https://arxiv.org/abs/1809.09603}{{\ttfamily 1809.09603}}].

\bibitem{Peng:2023aa}
H.~{Peng} and Y.~{Yu}, \emph{{Precise self-calibration of interloper bias in spectroscopic surveys}}, \href{https://doi.org/10.1093/mnras/stad2808}{\emph{Mon. Not. Roy. Astron. Soc.} {\bfseries 526} (2023) 820} [\href{https://arxiv.org/abs/2305.10487}{{\ttfamily 2305.10487}}].

\end{thebibliography}\endgroup






\end{document}